\documentclass[
superscriptaddress,
 amsmath,amssymb,
 aps,
pra,
twocolumn]{revtex4-2}
\usepackage{amsmath}
\usepackage{graphicx,stmaryrd}
\usepackage[utf8]{inputenc}
\usepackage[T1]{fontenc}

\usepackage{dcolumn}
\IfFileExists{newtxtext.sty}
   {\usepackage{newtxtext,newtxmath}}
   {\IfFileExists{stix.sty}
      {\usepackage{stix}}
      {\IfFileExists{mathptmx.sty}
      {\usepackage{mathptmx}}{} } }

\usepackage{textcomp}
\usepackage{multirow}

\usepackage{hyperref}

\usepackage{xcolor}
\definecolor{LinkColor}{rgb}{0.256,0.439,0.588}

\hypersetup{
   colorlinks=true,
   linkbordercolor = {white},
   citecolor=LinkColor,
   linkcolor={black},
   urlcolor=LinkColor
   }
\begin{document}


\title{Delocalized polaron and Burstein-Moss shift induced by Li in $\alpha$-$\textrm{V}_{2}\textrm{O}_{5}$: DFT+DMFT study}


\author{Huu T.\ Do$^\dagger$}
\affiliation{  Department of Chemical Engineering, University of Illinois Chicago, IL 60607, USA.}
\affiliation{ Materials Science Division, Argonne National Laboratory Argonne, IL 60439, USA.}
\thanks{Huu T.\ Do and Alex Taekyung Lee contributed equally to this work.}

\author{Alex Taekyung Lee$^\dagger$}
\affiliation{  Department of Chemical Engineering, University of Illinois Chicago, IL 60607, USA.}
\affiliation{ Materials Science Division, Argonne National Laboratory Argonne, IL 60439, USA.}

\author{Hyowon Park$^{2,}$}
\affiliation{Department of Physics, University of Illinois Chicago, IL 60607, USA.}

\author{Anh T. Ngo}

 \email{anhngo@uic.edu}
\affiliation{  Department of Chemical Engineering, University of Illinois Chicago, IL 60607, USA.}
\affiliation{ Materials Science Division, Argonne National Laboratory Argonne, IL 60439, USA.}

\date{\today}

\begin{abstract}

We performed density functional theory (DFT)+$U$ and dynamical mean field theory (DMFT) calculations 
with continuous time quantum Monte Carlo impurity solver to investigate the electronic properties of 
V$_2$O$_5$ and Li$_x$V$_2$O$_5$ ($x$ = 0.125 and 0.25). 
Pristine V$_2$O$_5$ is a charge-transfer insulator with strong O $p$-V $d$ hybridization, 
and exhibits a large band gap ($E_{\textrm{gap}}$) as well as non-zero conduction band (CB) gap.
We show that the band gap, the number of $d$ electrons of vanadium, $N_d$, and conduction band (CB) gap for V$_2$O$_5$ obtained from our DMFT calculations are in excellent agreement with the experimental values.  
While the DFT+$U$ approach replicates the experimental band gap, it overestimates the 
value of $N_d$ and underestimates the CB gap.
In the presence of low Li doping, the electronic properties of V$_2$O$_5$ are mainly 
driven by a polaronic mechanism, and electron spin resonance and electron nuclear double 
resonance spectroscopies observed the coexistence of free and bound polarons.
Notably, our DMFT results identify both polaron types, with the bound polaron being energetically preferred,
while DFT+$U$ method only predicts the free polaron. 
Our DMFT analysis also reveals that increased Li doping leads to electron filling in the conduction band, 
shifting the Fermi level, this result consistent with the observed Burstein-Moss shift upon enhanced Li doping, and 
we thus demonstrate that the DFT+DMFT approach can be used for accurate and realistic description of strongly correlated materials.

\end{abstract}

\maketitle

\section{\label{Introduction} Introduction}

Vanadium pentoxide (V$_2$O$_5$) is an interesting compound in the vanadium-oxide family,
since its highest oxidation state $+5$ ($d^0$) results in the strongest degree of an electronegativity 
and largest percentage of covalent bond in the oxide compounds,
and it can be easily reduced to  lower oxidation states
\cite{V2O5-review2023,Chem-Open}.
Furthermore, V$_2$O$_5$ is a long attractive exemplar for both the fundamental research of  electronic properties 
in  transition oxides \cite{Sipr-PRB-1999, Shin-PRB-1990, Horrocks-JPCC-2016, Wang-PRB-2016} 
and a variety of applications in photocatalysis, and smart window, 
especially in fabricating a cathode for electrochemical storage 
\cite{Meyer-JAP-2011, Simard-1955, Ruan-NatureCom-2022, Whittingham-1976, Lim-2013, Jesus-NatureCom-2016}.

\begin{figure}
 \centering
\includegraphics[scale=0.52]{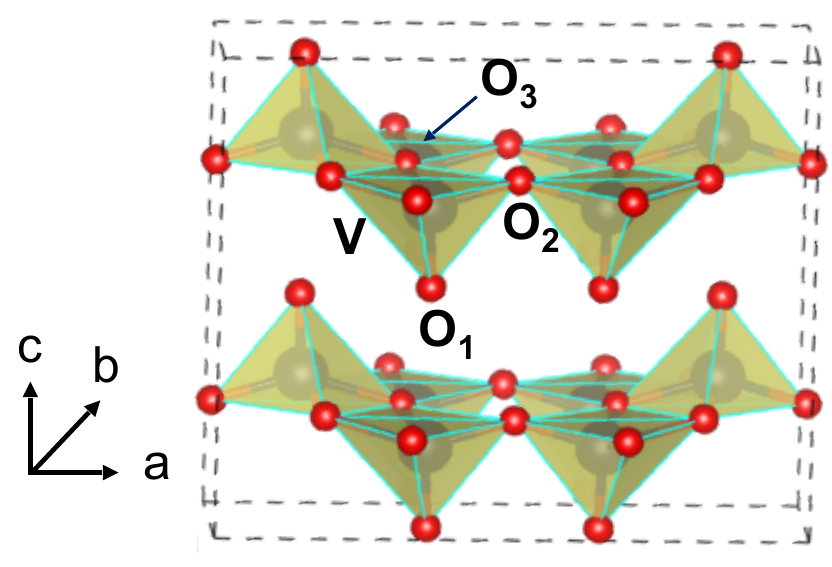}
\caption{Crystal structure of pristine $\alpha$-V$_2$O$_5$, which includes two layers 
(1$\times$2$\times$1 supercell). }
\label{figure1}
\end{figure}

V$_2$O$_5$ material exists in several polymorphs such as $\alpha$, $\beta$, $\gamma^{\prime}$ phases, 
\cite{Smirnove-IC-2018, Singh-2017}, and $\alpha$ phase is the most stable at ambient conditions.
$\alpha$-V$_2$O$_5$ has a layered structure (Figure~\ref{figure1}) with orthorhomic space group ($Pmmn$), 
and the layers interact with each other via a weak van der Waals force 
\cite{LONDERO20111805, Londero-PRB-2010, Sukrit-ACS-2017}.
$\alpha$-V$_2$O$_5$ is a charge-transfer insulator, and it has band gap of 2.3-2.8 eV
\cite{Kenny-1965, Choi-PRM-2019, Meyer-JAP-2011, Othonos_APL-2013}.
There is also a gap in the conduction band (CB gap) around 0.5 eV 
separating between the split-off band and the main conduction band \cite{Wang-PRB-2016}
(see Figure~\ref{figure2}). 
The number of electrons in the V $d$ manifold ($N_d$) was also measured experimentally.
Though V is in $d^0$ state, 
$N_d$ shows non-zero value due to the strong O $p$-V $d$ hybridization.
Resonant photoemission spectroscopy (RPES) estimated $N_d = 2.0$ for V$_2$O$_5$ \cite{WU2006309}, 
while cluster model predicted $N_d = 1.2$ based on XPS and X-ray absorption spectroscopies (XAS)
\cite{Mossanek-PRB-2008}.

Previous DFT+$U$ studies suggested the band gap of 1.5$-$2.2 eV, 
while the CB gap is only 0$-$0.15 eV
\cite{Scanlon-JPCC-2008,Jesus-NatureCom-2016,Roginski-JPCC-2021} . 
This value is much smaller than the experimental value of 0.4$-$0.5 eV.
Recent GW results showed that both band gap and CB gap are increased to 
2.4 eV and 0.3 eV, respectively.
However, to our knowledge, $N_d$ values are not reported in the past first-principles studies.
Since the $N_d$ represents the strength of the $p$-$d$ hybridization and because V$_2$O$_5$
is a charge-transfer insulator, $N_d$ plays a crucial role in deciphering the electronic properties of V$_2$O$_5$.

Due to the layer-by-layer structure, $\alpha$-V$_2$O$_5$ promises a potential candidate for the 
cathode material of Li-battery by intercalating $\mathrm{Li}^{+}$ ions between layers, in 
particularly for rechargeable microbatteries due to very high specific densities and capacities
\cite{Jarry-CM-2020,Zeng-2018}. 
Depending on the Li ratio, several phases of Li$_x$V$_2$O$_5$ are observed:
$\alpha$ ($x \leq 0.1$), $\epsilon$ ($0.33 \leq x  \leq 0.64$), or $\delta$ ($ 0.7 \leq x \leq 1.0 $) 
phases \cite{COCCIANTELLI1991103, Murphy-IC-1979, Hadjead-CM-2006}.

Li atoms in Li$_x$V$_2$O$_5$ donate electrons to V $d$ bands and becomes Li$^+$ ions.
With Li-doping, the electronic properties of $\alpha$-Li$_x$V$_2$O$_5$ 
are changed unexpectedly during the lithiation process.
If the Li concentration is low ($x$= 0.001 and 0.005), two types of polarons are observed experimentally:  
(i) free polarons localized at single V sites, 
and (ii) bound polarons delocalized over four V sites around a Li$^+$ ion 
\cite{Sanchez_1984,Pecquenard-1996}.
On the other hand, as $x$ increases, the optical band gap is increased with Li doping,
which indicates the Burstein-Moss shift, i.e., the Fermi level is shifted due to the doped 
electron in the conduction band \cite{Wang-PRB-2016}.

There are several DFT+$U$ studies of Li$_x$V$_2$O$_5$ in the literature, and they showed that 
the doped electron occupies the defect level located at the middle of the band gap,
while the conduction band is empty
\cite{Scanlon-JPCC-2008,Jesus-NatureCom-2016,Watthaison-2019,Suthirakun-JPCC2018}.
In these cases, the defect level or the electron is spatially localized on single V site,
similar to the free polaron, with the migration barrier of 0.12$-$0.34 eV.
\cite{Watthaison-2019,Suthirakun-JPCC2018}.
However, the bound polaron has not been found by DFT+$U$ studies. 
In addition, since the electron occupies the defect level, there is no shift of the Fermi level
within DFT+$U$.

In this work, we investigate the electronic structures of V$_2$O$_5$ and Li$_x$V$_2$O$_5$ 
($x$ = 0.125 and 0.25) using both DFT+$U$ and DFT+DMFT methods. 
We show that $N_d$ of V$_2$O$_5$ within DFT+$U$ approximates a twice of the experimental value, 
and $N_d$ using DMFT method is similar to experimental value.
We find that the bound polaron, where electron delocalized over four V sites, 
is energetically more stable than the free polaron within DMFT for Li$_{0.125}$V$_2$O$_5$.
Moreover, we show that as $x$ increases, the defect level is empty and electron occupies 
the conduction band within DMFT.
This result is consistent with the Fermi level shift or Burstein-Moss effect in the experiments 
\cite{Wang-PRB-2016, Jarry-CM-2020}.

We organize our manuscript as following. 
Sec.~\ref{methods} describes in details the optimization of crystal structure and electronic calculations using DFT+$U$ and DFT+DMFT.
In the Sec.~\ref{Results}, we  show and discuss atomic structure and electronic properties of pure V$_2$O$_5$ as well as   
their changes in  $\mathrm{Li}_x \mathrm{V}_2\mathrm{O}_5$ ($x = 0.125$ and 0.25).
We conclude the content of our paper in the Sec.~\ref{conclusion}.

\section{\label{methods} Computational details}

\subsection{\label{DFT} DFT+$U$ and Structural Optimization}

We performed density functional theory (DFT)+$U$ calculations with a combination of the rotationally invariant formalism 
and the fully localized limit double-counting formula \cite{LDA+U1} implementing inside VASP package
 \cite{DFT1993, DFT1994,DFT1996}.
 The projector augmented wave (PAW) method, which  describes the 
 relationship between core and valence electrons, was employed with the generalized gradient approximation (GGA) 
 of  Perdew-Burke-Ernzerhof (PBE) \cite{PBE1996}. 
 Since $\alpha$-V$_2$O$_5$ exhibits the layered structure (Figure~\ref{figure1}),
 van der Waals correction (vdW), specifically DFT-D2 method  \cite{Grimme-1, Grimme2}, was also applied to relax the structure. 

 \begin{figure}
 \centering
\includegraphics[scale=0.34]{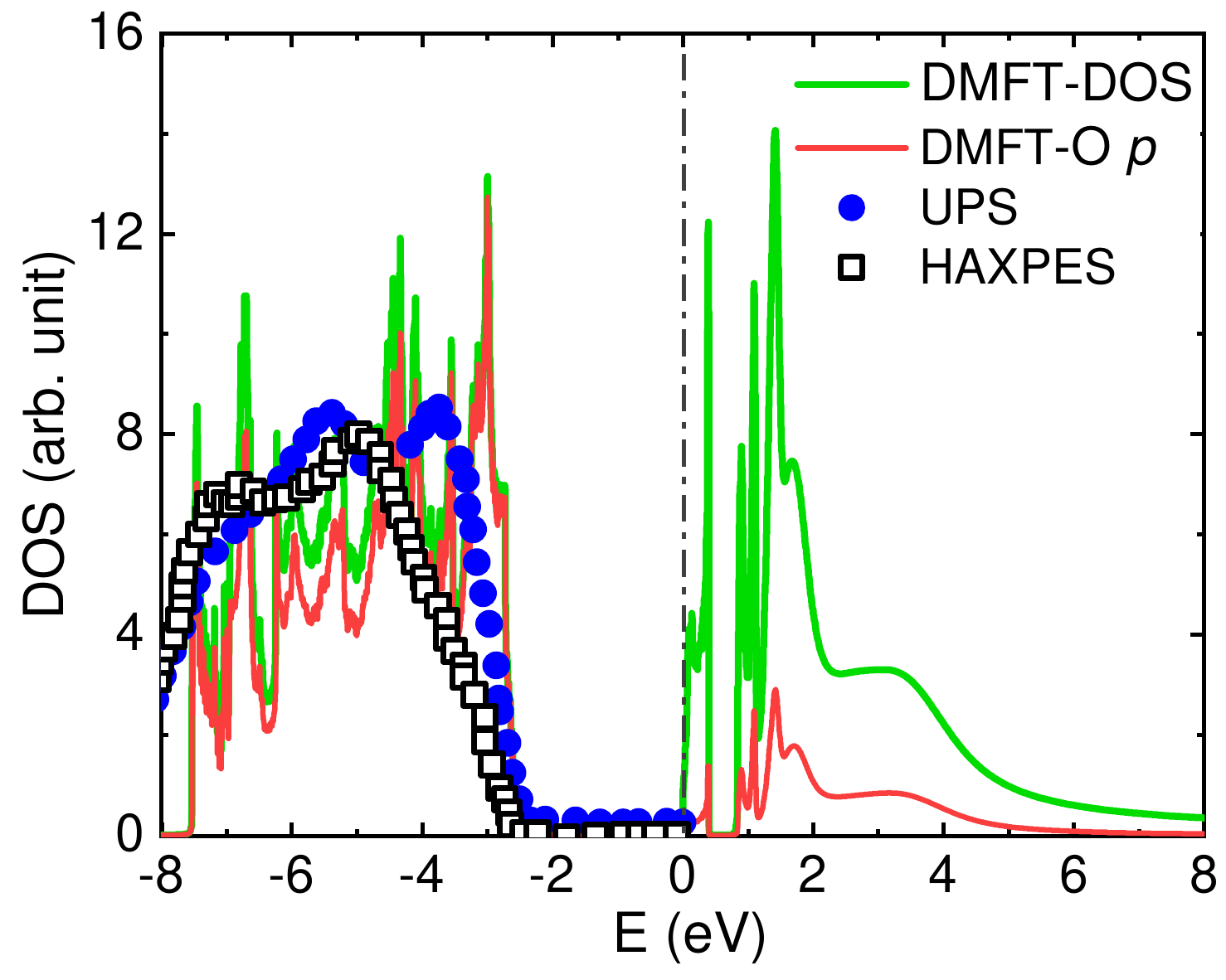}
\caption{DOS from experiments and DFT+DMFT. There is a gap in the conduction band (CB gap).
We use ultraviolet photoemission spectroscopy (UPS) data with photon energies of 32 eV in Ref.~\cite{Shin-PRB-1990}, 
and hard X-ray photoemission spectroscopy (HAXPES) data in Ref.~\cite{Jesus-NatureCom-2016}. Here, all experiments were measured at room temperature. 
The parameters for DMFT calculation are set by $U = 5.5$ eV, $J =0.5$ eV, $\lambda = 0.4$ and $T = 300$ K.   
 \label{figure2}
}  
\end{figure} 
For pristine and Li-doped V$_2$O$_5$, we adopted 1$\times$2$\times$1 and 1$\times$2$\times$2 supercells, 
that correspond to $\mathrm{V}_{8}\mathrm{O}_{20}$ 
and $\mathrm{V}_{16}\mathrm{O}_{40}$, respectively.
The Hubbard  $U$ parameter varied from 0 to 6 eV, while the Hund's coupling was fixed at $J = 0$ eV.
We utilized the kinetic energy cutoff of 600 eV, and $3\times9\times4$ and $3\times5\times3$ $k$-point meshes for 1$\times$2$\times$1 and 1$\times$2$\times$2 supercells, respectively.
The convergence of the structural relaxation was achieved once the atomic forces of all ions reached the value less than 0.01 eV/\AA{}.

\begin{table}[b]
\caption{ \label{tab:table0}
Lattice parameters of pristine $\alpha$-V$_2$O$_5$ using DFT+$U$ and DFT+$U$+vdW,
and experimental values. }
\begin{ruledtabular}
\begin{tabular}{lllll}
\textrm{$U$ (eV)} & \textrm{$a$ (\AA)} & \textrm{$b$ (\AA)} & \textrm{$c$ (\AA)} & \textrm{$V$ (\AA)$^3$ } \\
\colrule
0 & 11.558 & 3.562 & 4.729 & 194.706 \\
3 & 11.495 & 3.636 & 4.790 & 199.242 \\
4 & 11.485 & 3.636 & 4.790 & 199.901 \\
0+vdW & 11.634 & 3.532 & 4.427 & 181.919 \\
3+vdW & 11.563 & 3.588 & 4.460 & 185.030 \\
{\bf 4+vdW} & {\bf 11.548} & {\bf 3.606} & {\bf 4.469} & {\bf 186.127} \\
{\it Exp}.~\footnote{Experimental values are taken and averaged from X-ray data in Ref.~\cite{Smirnove-IC-2018, Singh-2017}.} 
&  11.548 & 3.577 & 4.392 & 181.454 \\
\end{tabular}
\end{ruledtabular}
\end{table}

We relaxed the lattice parameters of pristine $\alpha$-V$_2$O$_5$ 
using different $U$ values with and without vdW correction,
as summarized in Table~\ref{tab:table0}. 
The intercalating distance between two layers raises  
(parameter $c$ along $z$-direction in Figure~\ref{figure1}) 
with increasing $U$. 
Without vdW correction, $c$ parameter increases from 4.729 to 4.790 \AA~with  $U=0$ to 4 eV.
With vdW correction, $c$ parameter is suppressed from 4.427 to 4.469 \AA~with $U=0$ to 4 eV.
We conclude that DFT+$U$(=4 eV)+vdW gives the best lattice parameters compared 
to the experimental value $c = 4.392$ \AA~.

\begin{table}
\caption{\label{tab:table1}%
Lattice parameters of Li$_{0.125}$V$_2$O$_5$ and Li$_{0.25}$V$_2$O$_5$ 
using DFT+$U$ and DFT+$U$+vdW, and experimental data.
We relax the structures using $1\times2\times1$ and $1\times2\times2$ supercells, 
the values for the $1\times1\times1$ unit cell are shown.
}
\begin{ruledtabular}
\begin{tabular}{l l l l l l}
\textrm{x}&
\textrm{$U$ (eV)}&
\textrm{$a$ (\AA)}& \textrm{$b$ (\AA)}& \textrm{$c$ (\AA)}& \textrm{$V$ (\AA)$^3$ } \\
\colrule
\multirow{5}{*}{0.125} & 0 & 11.498 & 3.564 & 4.742 & 194.322 \\
 & 4 &  11.545 & 3.635 & 4.632 & 194.387 \\
 & 0+vdW & 11.562 & 3.534 & 4.453 & 181.941 \\
 & {\bf 4+vdW} & {\bf 11.565} & {\bf 3.604} & {\bf 4.444} & {\bf 185.227} \\
& {\it Exp}.~\footnote{Experimental values are taken and averaged from X-ray data in Ref.~\cite{Murphy-IC-1979, Talledo-JAP-1995}} & {\it 11.470} & {\it 3.570} & {\it 4.470} & {\it 183.037} \\
\colrule
\multirow{3}{*}{0.25} & 0+vdW & 11.521 & 3.538 & 4.444
 & 181.12 \\
  & {\bf 4+vdW} & {\bf 11.514} & {\bf 3.552} & {\bf 4.548} & {\bf 185.980} \\
 & {\it Exp}.~\footnote{Similar references to $x = 0.125$ case$^a$} & 11.410 & 3.570 & 4.540 & 184.931 \\
\end{tabular}
\end{ruledtabular}
\end{table}

The lattice parameters of Li$_{0.125}$V$_2$O$_5$ and Li$_{0.25}$V$_2$O$_5$ are listed in Table~\ref{tab:table1}.
The intercalating distance between two layers increases from 4.444 to  4.548 \AA~for $x$ = 0.125 and 0.25, respectively.
Similar to the pristine case, the DFT+$U$+vdW with  $U$ = 4 eV gives the best match with the experiments.
Since $U = 4$ eV also provides a reasonable band gap for the pristine $\alpha$-V$_2$O$_5$ 
(Figs.~\ref{figure3} and ~\ref{figure4}),
hereafter we focus on the structure obtained using $U$ = 4 eV within DFT+$U$+vdW, unless specified.

\subsection{DFT+DMFT method}
\label{DMFT_Method}

At the first step of a conventional DFT+DMFT procedure \cite{Park-prb-2020, Alex-prb-2021, Singh-CPC-2021}, 
we employ DFT+$U$+vdW to optimize atomic structures and subsequently create localized Wannier orbitals. 
For pristine V$_2$O$_5$, we use $U$ = 4 eV, as mentioned previously. 
On the other hand, in Li-doped V$_2$O$_5$, the addition of electrons to the system leads 
to the emergence of two types of polarons \cite{Sanchez_1984, Pecquenard-1996}, 
which result in distinct local structural distortions. 
However, the optimized structure obtained using $U$ = 4 eV only provides the free polaron, 
where the electron is localized at a single V site, and significant structural relaxation 
is confined to the vicinity of this electron-localized V site. 
Therefore, to capture the structural distortion induced by the delocalized (or bound) polaron, 
we also relax the structure with $U$ = 0 eV, while maintaining fixed lattice parameters. 
By examining these two different structures, we can compare the two types 
of polarons within both DFT+$U$ and DFT+DMFT.
Further details can be found in Appendix~\ref{detect_polarons}.

In the second step of DFT+DMFT calculations, V $d$ and O $p$ orbitals were constructed to represent 
a hybridization subspace by projecting the Kohn-Sham (KS) plane-wave functions onto 
maximally localized Wannier functions (MLWFs) \cite{MOSTOFI2008685}. 
In this step, non-spin polarized DFT ($U$ = 0 eV) scheme is used.
In the last step, V $d$ manifolds were implemented by using the continuous time 
quantum Monte Carlo (CTQMC) impurity solver within DMFT \cite{DMFT1, Park-prb-2020, Haule2015}. 
An additional unitary rotation transformation for the Wannier subspace of V $d$ orbitals was  
applied to minimize the off-diagonal hopping terms \cite{Park-prb-2020}.
In these systems, we consider the hybridized region within energy window of 10 eV around the Fermi level [see Fig.~\ref{appendix1} in Appendix~\ref{appendix:band}].

The rotationally invariant Coulomb interaction in the form of the Slater-Kanamori interaction Hamiltonian \cite{Slater-1951, Kanamori-1963, Kanamori-overview} is
\begin{equation}
\begin{split}
\hat{H}_\textrm{SK} =  \ &U \sum_{\alpha} \hat{n}_{\alpha\uparrow} \hat{n}_{\alpha\downarrow}
+\frac{1}{2}\sum_{\alpha \neq \beta} \sum_{\sigma \sigma'} \left( U' - J \delta_{\sigma \sigma'} \right) 
\hat{n}_{\alpha\sigma} \hat{n}_{\beta\sigma'} \\
 &-  \sum_{\alpha \neq \beta} \left( 
J c^{\dagger}_{\alpha\uparrow} c_{\alpha\downarrow} c^{\dagger}_{\beta\downarrow} c_{\beta\uparrow} +
J' c^{\dagger}_{\beta\uparrow} c^{\dagger}_{\beta\downarrow} c_{\alpha\uparrow} c_{\alpha\downarrow} 
\right). 
\end{split}
\label{sk-ham}
\end{equation}
Here, $c_{\sigma}$ and $c^{\dagger}_{\sigma}$ denote the fermion anihilation 
and creation operators, where $\sigma$ is spin.
$U$ denotes intra-orbital density-density interaction parameter, 
$U'$ is inter-orbital density-density interaction parameter,
$J$ is spin-flip interaction parameter, and 
$J'$ is pair-hopping interaction parameter.
$U'=U-2J$, $J'=J$ are due to the rotational invariance.

To investigate the temperature effect, we employed electronic temperatures of 300 K. 
This choice is motivated by the application of V$_2$O$_5$ as a cathode material for batteries, 
which typically operate at room temperature. 
For the single Li-doped V$_2$O$_5$, we also considered 150 K and found the electronic 
structures to be nearly indistinguishable. 
It is worth noting that within the CTQMC framework, we limited our considerations to 
density-density interactions. 
Given that there are 16 vanadium atoms in the supercell, a full Coulomb interaction 
calculation would be computationally demanding.

In the DMFT self-consitent calculations, the convergence of self-energy is determined once local or lattice self-energy $\Sigma^{\textrm{loc}} ( i \omega_n)$ approaches the impurity self-energy $\Sigma^{\textrm{imp}} (i \omega_n)$, with the discrete Matsubara frequency $\omega_n$ \cite{DMFT1, Park-prb-2014}. \{Note that the self-energy  is approximated as a local quantity in the correlated subspace, i.e., $\Sigma (\mathbf{k}, i\omega_n) \simeq  \Sigma(i\omega_n)$ \cite{DMFT1}.\}
So, the total DFT+DMFT energy is given by:
\begin{equation}
 \label{E_total1}
    E^{\textrm{TOT}} = E^{\textrm{DFT}} (\rho) + \sum_{m, \mathbf{k}} \epsilon_{m}(\mathbf{k}) \cdot \big [ n_{mm}(\mathbf{k}) - f_{m}(\mathbf{k}) \big ]
    + E^{\textrm{POT}} - E^{\textrm{DC}},
\end{equation}
where $E^{\textrm{DFT}}$ is the DFT energy computed by the electronic charge density $\rho$. 
$\epsilon_{m}(\mathbf{k})$ denotes as the DFT eigenvalues, and $n_{mm}(\mathbf{k})$ and $f_{m}(\mathbf{k})$ are the diagonal DMFT occupancy matrix element and Fermi function, respectively, with the KS band $m$ and momentum $\mathbf{k}$.
The potential energy $E^{\textrm{POT}}$ is calculated by using Migdal-Galiski formula: \cite{Migdal-1958}:
\begin{equation}
    E^{\textrm{POT}} = \frac{1}{2}  \sum_{\omega_n} \big [ \Sigma^{\textrm{loc}}(i\omega_n) \cdot G^{\textrm{loc}} (i\omega_n)].
\end{equation}
Here, the local Green's function is simplified by $G^{\textrm{loc}} (i\omega_n) = \sum_{\mathbf{k}} G^{\textrm{loc}} (\mathbf{k},i\omega_n)$ \cite{DMFT1, Park-prb-2014}.

Similar to the conventional fully localized limit, we used a double counting energy $E^{\textrm{DC}}$ \cite{Singh-CPC-2021,Park-prb-2014, Alex-prb-2021} to consider to the double counting corrections for DFT+DMFT calculation as: 
\begin{equation}
    E^{\textrm{DC}} = \frac{(U-\lambda)}{2} N_d \cdot (N_d-1) - \frac{J}{4} N_d \cdot (N_d-2), 
\end{equation}
where $N_d$ is called the formal $d$-electron number obtained self-consistently at each V $d$ site,
and $\lambda$ is the double counting parameter \cite{Singh-CPC-2021}.
$N_d$ is directly computed from the local Green function $G^{\textrm{loc}}(\mathbf{k}, \mathbf{k}^{\prime},i \omega_n)$:
\begin{equation}
\label{Nd1}
    N_d = \sum_{a, n} \sum_{ \mathbf{k}, \mathbf{k}^{\prime}}  \textrm{Im} \big \{  [ \phi^a_d(\mathbf{k})]^{\ast} G^{\textrm{loc}}(\mathbf{k}, \mathbf{k}^{\prime}, i \omega_n) \phi^{a}_d (\mathbf{k}^{\prime}) \big \}.
\end{equation}
Here, the $\omega_n$ is the Matsubara  frequency, and $\phi^a_d(\mathbf{k})$ represents  
the normalized $d$-orbital wave-function, which  is transformed from $\phi^a_d(\mathbf{r})$ with the real coordinates $\mathbf{r}$ positioning on a transition metal ion \cite{Xin-prb-2012}.
The spectral function or density of state (DOS) is calculated by  using the maximum entropy method \cite{MaximumEntro-1989}: 
\begin{equation}
    A(\omega_n) = -\frac{1}{\pi} \textrm{Im} \Big [ \sum_{\mathbf{k}} G^{\textrm{loc}}(\mathbf{k}, \omega_n)\Big].
\end{equation}

\section{\label{Results} Results and Discussion}

\subsection{\label{sec:V2O5} Pristine $\alpha$-V$_2$O$_5$ }

As mentioned, $\alpha$-V$_2$O$_5$ has layered structure with van der Waal interaction between the layers.
A vanadium atom stands at a distorted pyramidal coordination surrounding by five oxygen atoms 
which are classified into three different types, as depicted in Figure~\ref{figure1}: 
(i) vanadyl oxygen ($\mathrm{O}_{1}$ forms a double bond with the vanadium atom), 
(ii) bridge oxygen ($\mathrm{O}_{2}$ connects two vanadium atoms in different chain) and 
(iii) chain oxygen ($\mathrm{O}_{3}$ bonds to three vanadium atoms) \cite{Sipr-PRB-1999, Eyert-PRB-1998}.

\begin{figure}
 \centering
\includegraphics[scale=0.22]{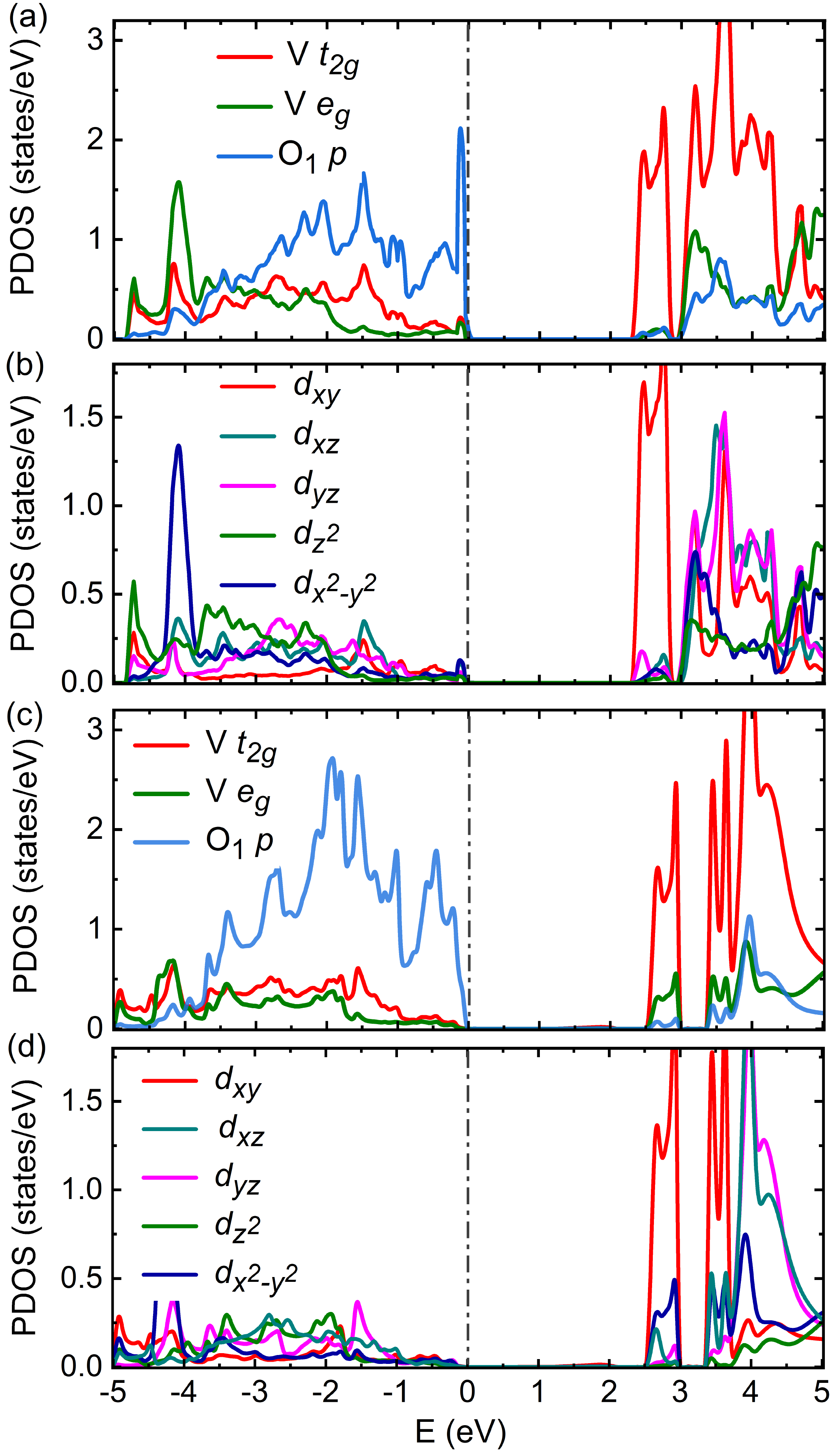}
\caption{ Projected density of states onto V $d$ and O $p$ of pristine V$_2$O$_5$,
using (a)-(b) DFT+$U$ with $U = 4$ eV and $J = 0$ eV,
and (c)-(d) using DFT+DMFT with $U = 5.5$ eV, $J = 0.5$ eV. 
$\lambda = 0.4$ and temperature $T = 300$ K are used.
Valence band maximum (VBM) is set to be zero.
 \label{figure3}}   
\end{figure}

V has 5+ charge state with $d^0$ in V$_2$O$_5$, and thus the conduction bands 
are largely dominated by V $d$ bands, 
whereas the valence bands near the Fermi level are significantly from O $p$ bands
(Figures~\ref{figure2} and \ref{figure3}).
From the structure of V-O bonds, V $d_{x^2-y^2}$ and $d_{z^2}$ forms $\sigma$ bonds 
$p$ orbitals of O$_1$ and O$_2$+O$_3$ atoms, respectively,
while $t_{2g}$ orbitals forms $\pi$ bonds with O atoms. 
Since one of the apical oxygen is missing compared to the VO$_6$ octahedron,
the cubic symmetry of $d$ bands is broken. 
Thus, doubly degenerate $e_g$ bands split into $d_{x^2-y^2}$ and $d_{z^2}$ bands, and 
$t_{2g}$ bands break into $d_{xy}$ band as well as double degeneracy of $d_{xz}$+$d_{yz}$ 
bands, as presented in Figure \ref{figure3}.
As a result, there is splitting in the V $d$ states in the conduction band, 
as presented in Figure~\ref{figure3}.
The lower band called as "split-off band", and the higher band is named as "main conduction band".

\begin{figure}
 \centering
\includegraphics[scale=0.33]{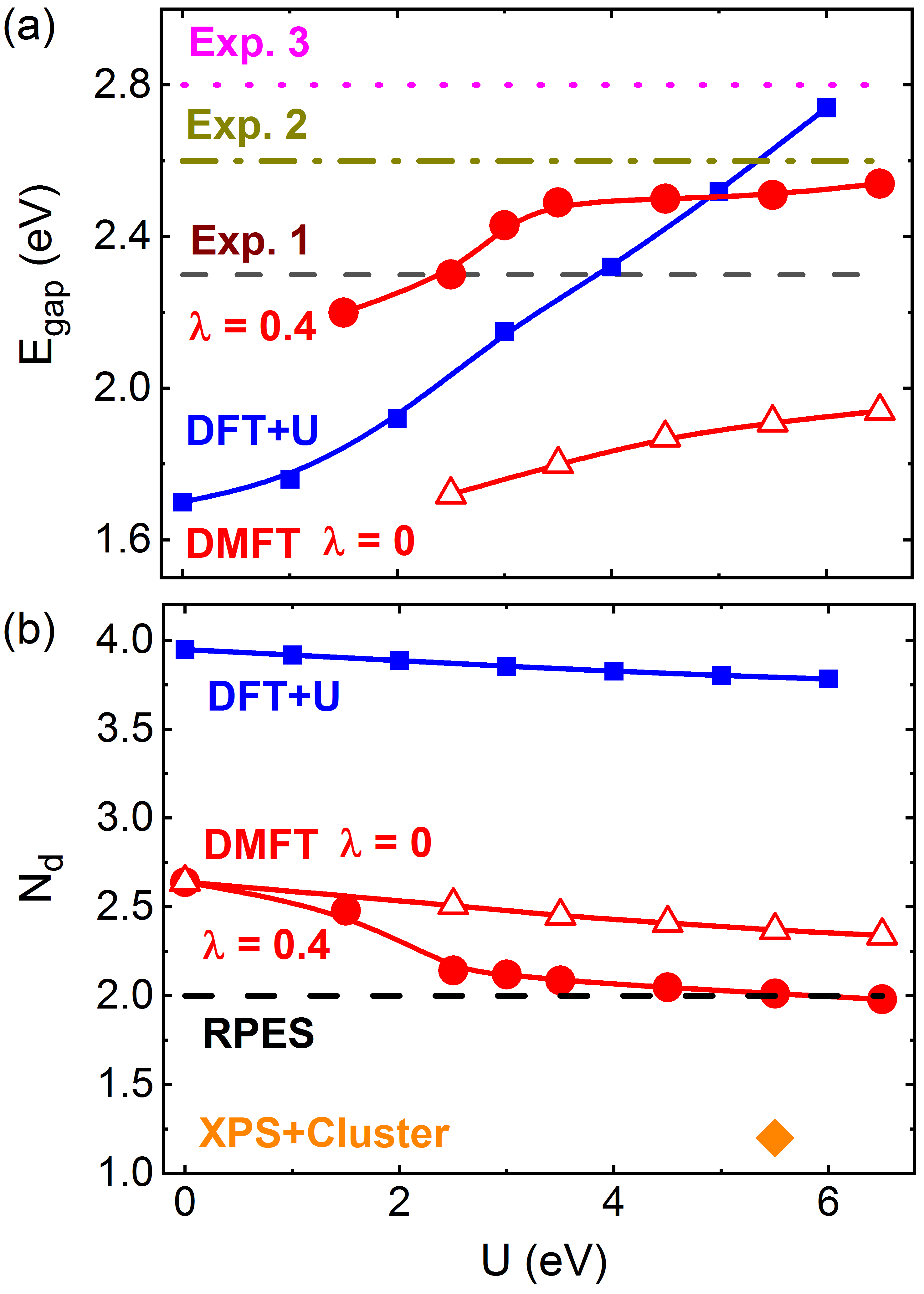} 
\caption{(a) Energy gap ($E_\textrm{gap}$) of pristine V$_2$O$_5$ as a function of $U$. 
The experimental values of the band gap are taken from Ref.~\cite{Kenny-1965} (Exp.~1, black dashed line), Ref.~\cite{Choi-PRM-2019} (Exp.~2, dark yellow dashed dotted line) and  
Ref.~\cite{Meyer-JAP-2011} (Exp.~3,  magenta dotted line).
(b) the number of $d$ electron ($N_d$) of V atom in V$_2$O$_5$, as a function of $U$.
Data of RPES and cluster model are from Ref.~\cite{WU2006309} and 
Ref.~\cite{Mossanek-PRB-2008}, respectively. 
$J$ = 0 is used for DFT+$U$, and $J =0.5$ eV and $T = 300$ K are used for DMFT calculations.
Two different values of the double counting parameter are shown: $\lambda$ = 0 and 0.4.
\label{figure4}}
\end{figure}

 We first study the effect of $U$ on the width of the energy gap ($E_\mathrm{gap}$), and the CB gap  
(due to separating between the split-off band and the main conduction band) using DFT+$U$, 
as illustrated in Figure~\ref{figure4}(a).
At $U = 0$ eV, $E_{\textrm{gap}}$ = 1.7 eV is much smaller than the experimental values 
of 2.3$-$2.8 eV
\cite{Kenny-1965,Choi-PRM-2019,Meyer-JAP-2011}. The CB gap of 0.4 eV is comparable to the experimental splitting 0.5 eV \cite{Wang-PRB-2016, Le-2018}.
At $U$ = 4 eV, $E_\textrm{gap} = 2.3$ eV agrees well with prior DFT+$U$ studies 
\cite{Mossanek-PRB-2008, Smirnove-IC-2018} and the experiments.
However, the CB gap is only 0.20 eV, which is narrower than the experimental result \cite{Wang-PRB-2016, Le-2018}. Therefore, we remark that $E_\textrm{gap}$ ascends with respect to $U$, whereas 
the CB splitting width descends versus $U$.

In Fig.~\ref{figure4}(b),  we observe that DFT+$U$  overestimates the number of $d$ electron  ($N_d$) in V. 
This value is around 3.8$-$4.0 in the $U$ range of  0$-$6 eV, but  
 larger than the experimental ones such as 
 $N_d = 2.0$ measured by RPES   \cite{WU2006309},
and 1.2 calculated by the cluster model based on XPS and XAS 
\cite{Mossanek-PRB-2008}. 
Therefore, DFT+$U$ has a critical limitation to describe the physics of V $d$ bands.
We also note that the $N_d$ value depends on the projection methods.
At $U$ = 0 eV, $N_d$ using the Wannier projectors gets 2.64 
[corresponds to DMFT with $U=0$ in Figure~\ref{figure4}(b)], 
and it is smaller than $N_d$ = 4.0 from the PAW projectors.

In order to resolve the limitation of DFT+$U$, we performed DFT+DMFT calculations for pristine $\alpha$-V$_2$O$_5$.
First, we compute $E_\textrm{gap}$ as a function of $U$ without the double counting parameter (i.e., $\lambda=0$),
 [Figure~\ref{figure4}(a)]. At $\lambda=0$, $E_\textrm{gap}$ is 1.71$-$1.95 eV with $U$ = 2.5$-$6.5 eV and less sensitive on $U$ in comparison with DFT+$U$,
and it does not approach the experimental value of 2.3$-$2.8 eV
\cite{Kenny-1965,Choi-PRM-2019,Meyer-JAP-2011}.
Since V$_2$O$_5$ exhibits as a charge-transfer insulating system, the $p$-$d$ hybridization is more important
than the $d$-$d$ correlation for determining  its band gap. 
The double counting correction $\lambda$  controls the degree of $p-d$ covalency, increasing this parameter results in larger separation between O $p$
and V $d$ bands and thus enlarges $E_\textrm{gap}$.
At $\lambda=0.4$, we  obtain $E_\textrm{gap}$= 2.35$-$2.55 eV for U values of 2.0$-$6.5 eV, which are in very good agreement to the experimental values.

Nonzero $\lambda$ is also needed for the reasonable $N_d$, 
as depicted in
Figure~\ref{figure4}(b).
The $N_d$ values using $\lambda=0$ with $U$ = 1.5$-$6.5 eV are  2.50$-$2.34, 
always larger than the experimental values
\cite{WU2006309,Bocquet-PRB-1996, Mossanek-PRB-2008}.
If $\lambda=0.4$ is  implemented, we show $N_d$= 2.50$-$1.98 with the range of $U$ from 2.5 to 6.5 eV. Particularly, $N_d$ = 2.01 for $U$ = 5.5 eV matches well with the RPES value  \cite{WU2006309}.
Therefore, combining the results of $E_\textrm{gap}$ and $N_d$, we conclude that $U$ = 5.5 eV and $\lambda$ = 0.4
are the best parameters for DMFT computations, with $E_\textrm{gap}$ = 2.52 eV and  $N_d$ = 2.01.

\begin{table}
\caption{ Contribution of $3d^{n}L^n$ states to the ground state of V$_2$O$_5$. 
\label{table:ground_state}}
\begin{ruledtabular}
\begin{tabular}{l| ll}
Configuration & DMFT & cluster \cite{Mossanek-PRB-2008}   \\
\hline
$3d^{0}$ & 2\% & 20\%  \\
$3d^{1}L$ & 22\% & 47\%   \\
$3d^{2}L^2$ & 41\% & 28\%   \\
$3d^{3}L^3$ & 25\% & N/A   \\
\end{tabular}
\end{ruledtabular}
\end{table}

Different from our DMFT calculation and the previous RPES measurement ($N_d = 2.0$) 
\cite{WU2006309}, Mossanek \emph{et al.} showed $N_d \sim 1.2$ from the 
single impurity cluster model, solved by configuration interaction method \cite{Mossanek-PRB-2008}.
They considered $\left[\mathrm{VO}_5 \right]^{-5}$ (V$^{5+}$)  
with $C_{4v}$ symmetry, corresponds to the square base pyramid structure.
In order to explain the difference of $N_d$ between  our DMFT and the cluster model, 
we calculate the contribution of $d^{n}L^n$ configurations
to the ground state of $\alpha$-V$_2$O$_5$, as summarized in Table \ref{table:ground_state}. 
 The weight of $d^{0}$ is only 2\% within DMFT, 
whereas it is 20\% from the  cluster model \cite{Mossanek-PRB-2008}. 
From our DMFT calculations, $3d^{2}L^2$ configuration has the largest weight of 41\%,
and $3d^{1}L^1$ and $3d^{3}L^3$ account for 22\% and 25\% of population probabilities, respectively.
Within the cluster model, $3d^{1}L^1$ contributes the largest probability of 47\%,
and the weight of $3d^{2}L^2$ configuration is 28\% \cite{Mossanek-PRB-2008}.
 We note that the single impurity cluster model does not include the hybridization
between clusters, i.e., there are no V-V nor O-O interactions, and therefore
both $d$ and $p$ state do not have dispersion.
The absence of the band dispersion in their model may give rise to the suppression of $N_d$.

Spectral functions, i.e., DOS from DMFT calculations using $U$ = 5.5 eV and $\lambda$ = 0.4 
are presented in 
Figures~\ref{figure3}(c) and (d).
The CB gap of 0.4 eV between the split-off band and the main conduction band is close to the experimental value 0.5 eV from the photolumninesense measurements 
\cite{Wang-PRB-2016, Le-2018}.
We emphasize that while DFT+$U$ only provides reasonable value of $E_\textrm{gap}$,
experimental values of $E_\textrm{gap}$, $N_d$ and CB gap are successfully reproduced by  our DMFT calculations. That 
implies the accurate method for the electron correlation is essential, 
even for $d^0$ band systems.
Similar to DFT+$U$, the split-off band is mainly from $d_{xy}$ band within DMFT.
The O $p$ character is dominant near the valence band maximum,
especially between $-$3.5 to $-$2.0 eV [Figure~\ref{figure3}(c)].
%
In figure~\ref{figure2}, we also show that our DMFT DOS are well matched to UPS  \cite{Goodman-prb-1994}
 and HAXPES \cite{Jesus-NatureCom-2016} experiments, particularly at 
the positions of the Fermi level and the range of valence band.

\subsection{  $\alpha$-Li$_{x}$V$_2$O$_5$ ($x$ = 0.125 and 0.25) }
\label{sec:LixV2O5}
 In this section, we consider Li-doped $\alpha$-V$_2$O$_5$, including  
$\alpha$-Li$_{0.125}$V$_2$O$_5$ and $\alpha$-Li$_{0.25}$V$_2$O$_5$,
using both DFT+$U$ and DMFT methods.

\subsubsection{ $\alpha$-Li$_{0.125}$V$_2$O$_5$}
\label{sec:Li0.125V2O5}

\begin{figure*}
 \centering
\includegraphics[width=0.70\textwidth, angle=0]{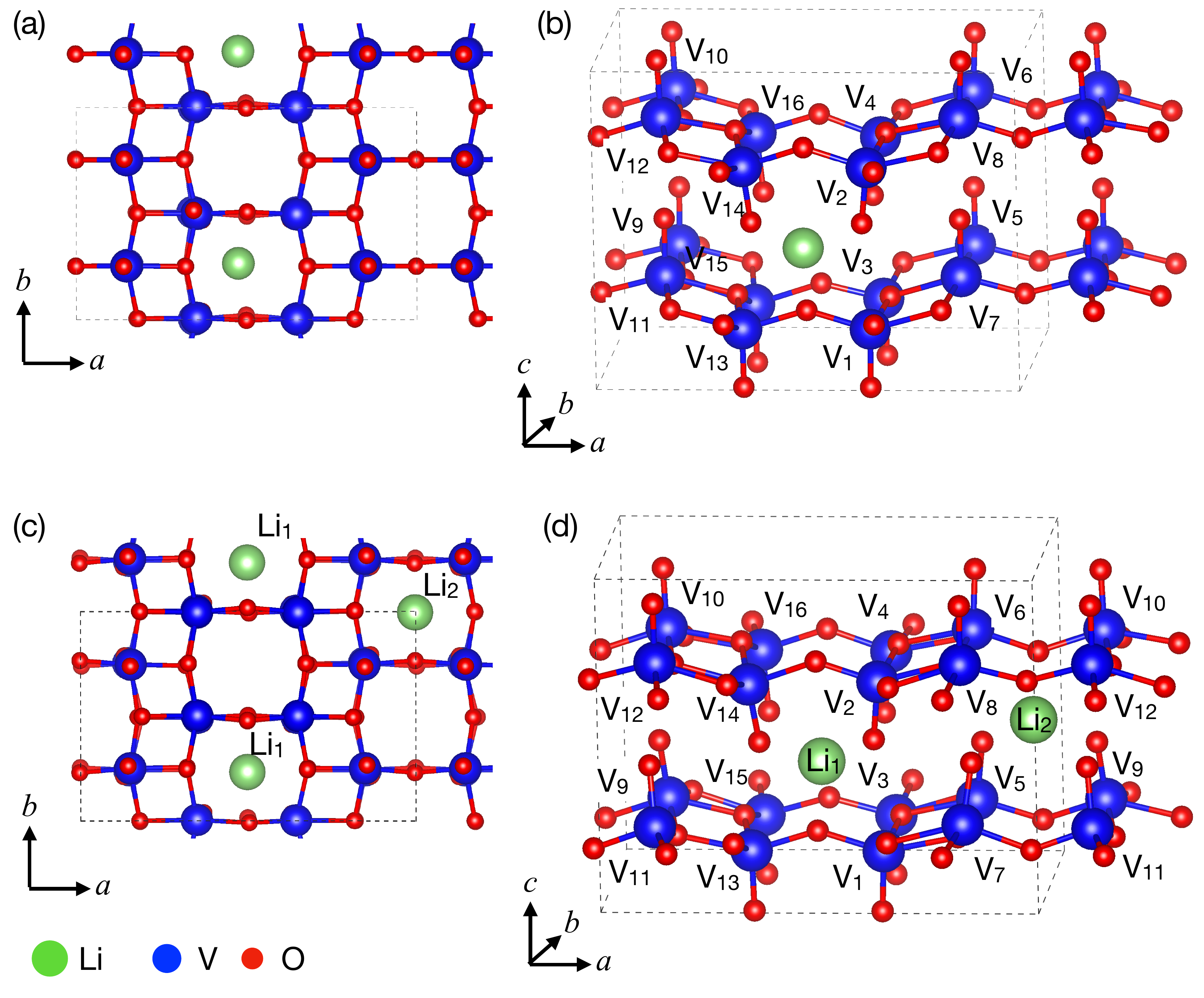}
\caption{ Atomic structure of (a)-(b) Li$_{0.125}$V$_2$O$_5$ and (c)-(d) Li$_{0.25}$V$_2$O$_5$.  
Indices of V atoms are shown. 
 \label{fig:LixV2O5_atmstr}}  
\end{figure*}

 $\alpha$-Li$_{0.125}$V$_2$O$_5$ is formed by intercalating one Li atom in the 
1$\times$2$\times$2 supercell (corresponds to Li$_1$V$_{16}$O$_{40}$).
The distance between Li ion and another one in the next supercell is 11.55 \AA, 7.21 \AA, and 8.94 \AA 
~along $a$, $b$, $c$ directions, respectively.
Thus, we assume that the interaction between Li defects are almost negligible once the periodic boundary condition is implemented.
We examined several initial different positions of doping single Li ion 
(see Appendix \ref{appendix:Li1}), 
and found that the most stable position of Li ion is middle of a hole,
as depicted in 
Figure \ref{fig:LixV2O5_atmstr}(a).
Li atom gets closer to the lower layer than the upper layer [Fig.~\ref{fig:LixV2O5_atmstr}(b)].
Distances from Li ion to the lower and upper V layers are 3.506 
and  5.381 \AA, respectively.
The stable position of Li ion is similar to structure obtained in the previous DFT+$U$ studies 
\cite{Scanlon-JPCC-2008, Jesus-NatureCom-2016}.

\begin{figure}
\centering
\includegraphics[width=0.40\textwidth, angle=0]{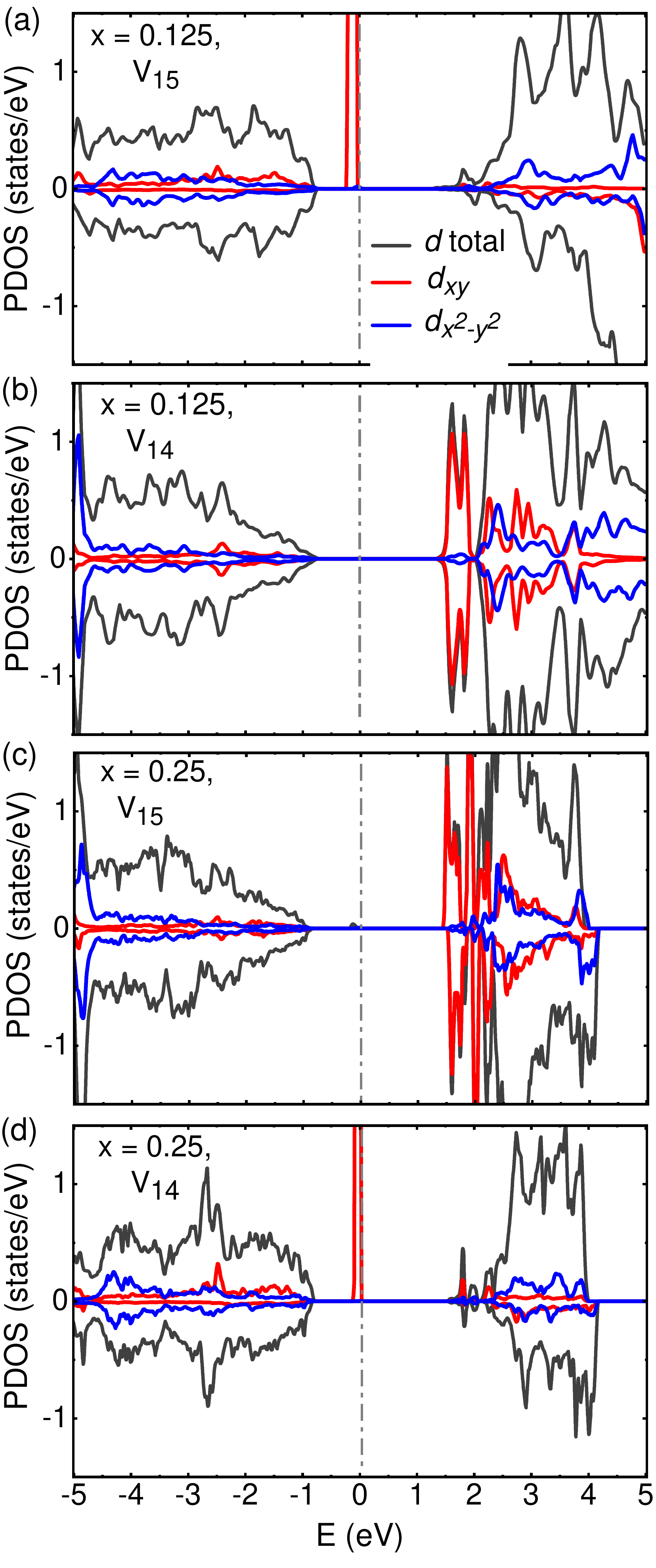}
\caption{DOS and projected density of states (PDOS) onto V $d$ orbital for 
(a)-(b) $\alpha$-Li$_{0.125}$V$_2$O$_5$ and  
(c)-(d) $\alpha$-Li$_{0.25}$V$_2$O$_5$, using DFT+$U$.
$U = 4$ eV and $J = 0$ eV are used. 
Fermi energy is set to be zero.
\label{fig:1Li-DFT-dos} }   
\end{figure}

\begin{table}[]
\caption{\label{table:LixV2O5-dft}
Magnetic moments ($\mu_{B}$) of V atoms in $\alpha$-Li$_{x}$V$_2$O$_5$ 
( $x = 0.125$ and 0.25), within DFT+$U$ calculations.
$U = 4$ eV and $J = 0$  are used. 
Indices of V atoms are shown in Figure~\ref{fig:LixV2O5_atmstr}. }
\begin{ruledtabular}
\begin{tabular}{l l l l l l l l l l}
$x = $ & $\mathrm{V}_1$ & $\mathrm{V}_2$ &
$\mathrm{V}_3$ & $\mathrm{V}_7$ &
$\mathrm{V}_{11}$ & $\mathrm{V}_{13}$ &
$\mathrm{V}_{14}$ & $\mathrm{V}_{15}$ & other V  \\
\colrule
0.125 & 0.01 & 0.00 & 0.08 & 0.00 & 0.00 & 0.01 & 0.02 & 1.11 & 0.00 \\
0.25 & 0.00 & 0.13 & 0.00 & 1.08 & 0.12 & 0.00 & 1.08 & 0.00 & 0.00
\end{tabular}
\end{ruledtabular}
\end{table}

 Once Li atoms are doped, they donate one electron per Li ion to V$_2$O$_5$ system, and become Li$^+$.
Li $s$ bands are fully empty and far above the Fermi level by 6.5 eV, indicating Li$^+$.
Within DFT+$U$, the splitting between the split-off band and the main CB becomes 
even smaller for Li$_{0.125}$V$_2$O$_5$, and the CB gap is nearly zero, while the CB gap is 
0.1 eV for pristine $\alpha$-V$_2$O$_5$ within DFT+$U$ 
(Figure \ref{fig:1Li-DFT-dos}).

 DFT+$U$ results point out 
a defect level is created at middle of the band gap for doping a single Li-ion in V$_2$O$_5$ framework [Figs.~\ref{fig:1Li-DFT-dos}(a)-(b)].
The defect band occupies one electron,
spin-up defect level is filled and located at 0.62 eV above VBM,
while spin-down level is empty.
The position of the spin-up defect level is similar to the previous DFT+$U$(=4 eV) study of 
$\alpha$-Li$_{0.028}$V$_2$O$_5$ (correspond to Li$_{1}$V$_{72}$O$_{180}$),
where Li defect level is near VBM+1.0 eV \cite{Scanlon-JPCC-2008}.

The origin of the defect level is the charge disproportionation of the V atoms,
since the electron occupying the defect level spatially localizes on single V atom.
Given that only spin-up defect band is occupied, magnetic moments of V atoms shown in 
Table \ref{table:LixV2O5-dft} directly indicates the charge disproportion in 
Li-doped V$_2$O$_5$ within DFT+$U$. 
Specifically, one electron from Li is donated at V$_{15}$ atom, which is the nearest neighbor
of Li atom with distance of 3.075 \AA~(see Figure \ref{fig:LixV2O5_atmstr}).





The localization of the electron  induces a polaronic effect in Li$_{x}$V$_2$O$_5$ 
\cite{Pecquenard-1996,Sanchez_1984, Jesus-NatureCom-2016, Ioffe-PSS1970}.
The ESR and ENDOR spectroscopies, and electronic conductivity measurement proposed  
two types of charge carriers in Li$_{0.005}$V$_2$O$_5$ \cite{Pecquenard-1996} 
and Li$_{0.001}$V$_2$O$_5$ \cite{Sanchez_1984}:
(i) free polarons localized at single vanadium sites, and
(ii) bound polarons delocalized over four vanadium sites around a Li$^+$ ion
(see Fig.~\ref{fig:Polaron} for schematic illustrations).
Since the electron prefers to occupy a single V, this corresponds the free polaron [Fig.~\ref{fig:Polaron}(a)].  
According to 
previous DFT+$U$ studies, the electron could be positioned on other V sites with higher value than the ground state by 0.1$-$0.2 eV  \cite{Watthaison-2019}.
The migration barrier from DFT+$U$ calculations are 0.12$-$0.34 eV
\cite{Watthaison-2019,Suthirakun-JPCC2018},
close to the experimental values 0.27$-$0.28 eV \cite{Ioffe-PSS1970,Giannetta-2015}.
However, the bound polaron 
has not been observed by DFT+U calculations. 

\begin{table}[b]
        \caption{Energy difference between  free and bound polarons calculated within DFT+$U$ and DMFT  
        for Li$_{0.125}$V$_{2}$O$_{5}$.
        Here, we used spin-polarized DFT+$U$ with $U = 4$ eV and $J = 0$ eV. The parameters for DMFT are $U = 5.5$ eV and $J = 0.5$ eV, $\lambda=0.4$ at room temperature.
     We set the energy level  of the free polaron at  0 eV and compare with bound one.            \label{tab:polaron}}
    \begin{ruledtabular}
    \begin{tabular}{c c c}
         Methods & free polaron & bound polaron \\
         \hline
         DFT+U & 0
 &    0.51 eV
         \\
         DMFT & 0 &   $-$0.11 eV
    \end{tabular}
\end{ruledtabular}
\end{table}

\begin{figure}
 \centering
\includegraphics[scale=0.30]{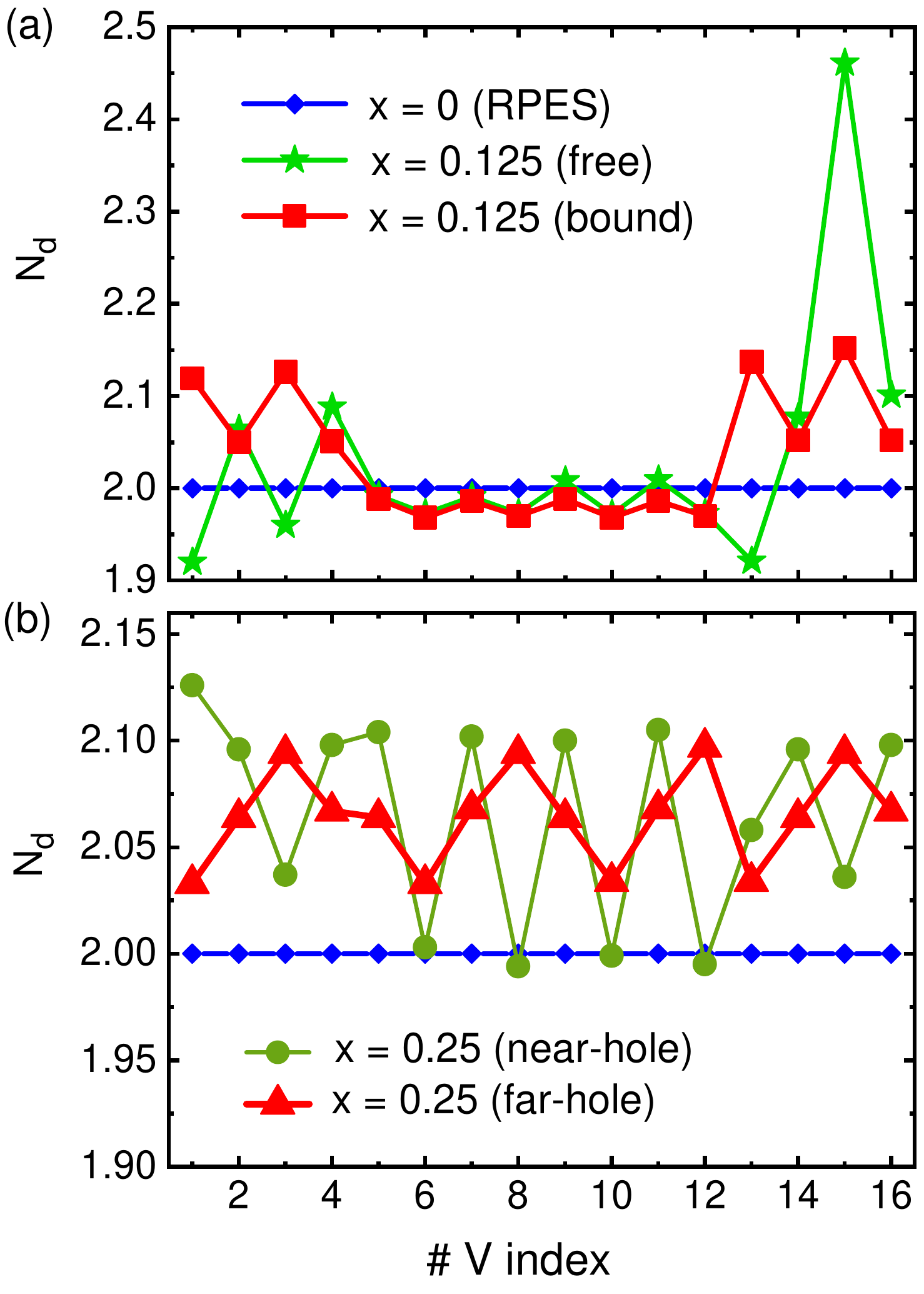}
\caption{  $N_d$ of V atoms in $\mathrm{Li}_x \mathrm{V}_{2}\mathrm{O}_{5}$ for (a) $x = 0.125$ (b) $x = 0.25$.  
The $N_d$ value is calculated by DFT+DMFT method with $U = 5.5$ eV, $J = 0.5$ eV, $\lambda=0.4$ at $T = 300$ K. 
We consider $\mathrm{Li}^+$ ion is the center of system, we classify the $N_d$ values into high $N_d$ value for nearest V atom, medium value for next-nearest and low value for far V atoms [see Fig.~\ref{fig:LixV2O5_atmstr}]. 
Blue diamond with line show the background homogeneous $N_d$ of V atom in our pristine V$_2$O$_5$ and RPES measurement \cite{WU2006309}. 
For which V atoms have $N_d > 2.0$ will receive a electron donated by Li atom. 
\label{fig:LixV2O5-Nd}
}
\end{figure}

\begin{figure}
 \centering
\includegraphics[width=0.35\textwidth]{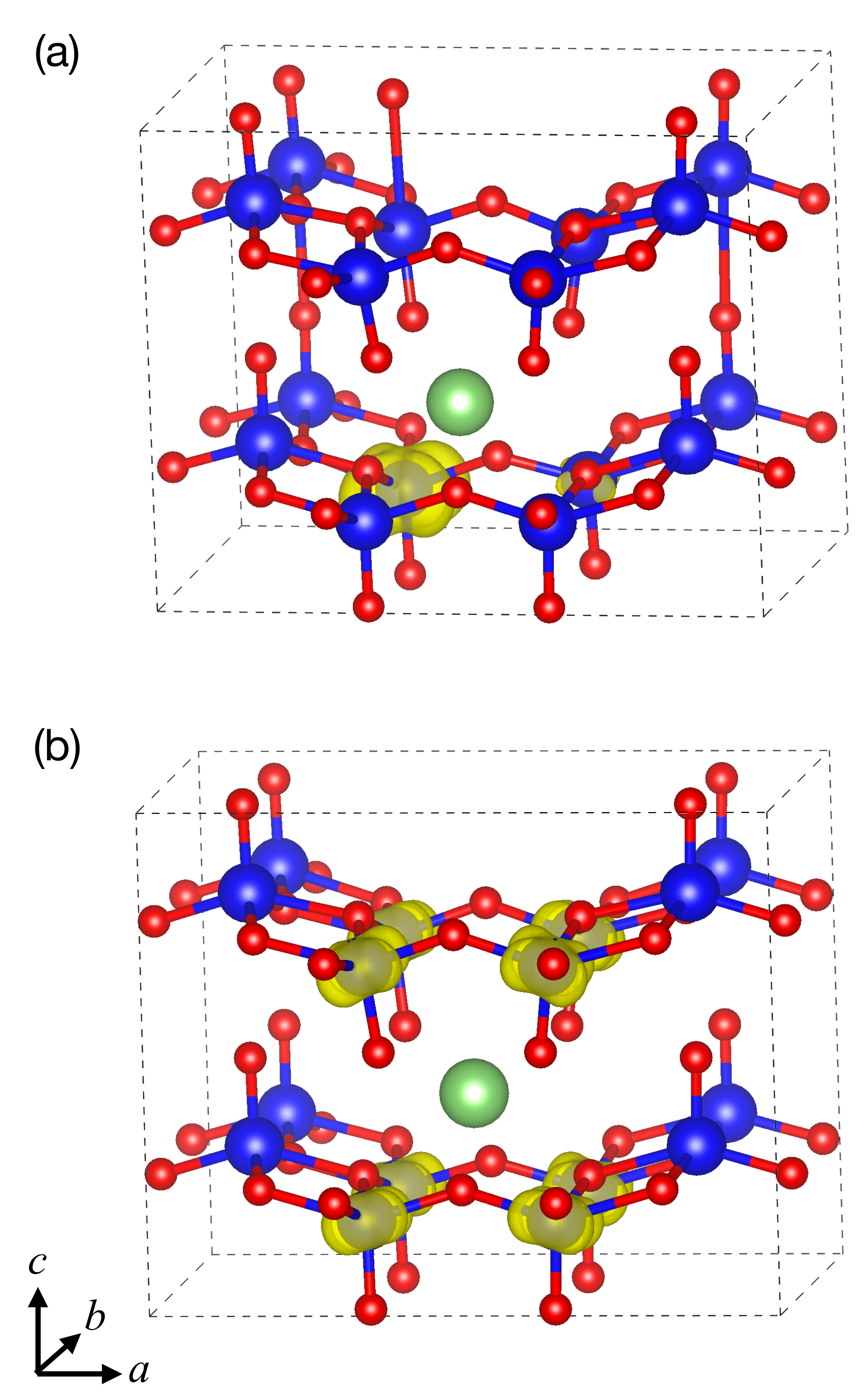}
\caption{ Schematic illustration of (a) free and (b) bound polarons in Li$_{0.125}$V$_2$O$_5$. 
The atomic positions are referred to Fig.~\ref{fig:LixV2O5_atmstr}(b) . 
\label{fig:Polaron}
}
\end{figure}

\begin{figure*}
 \centering
\includegraphics[scale=0.227]{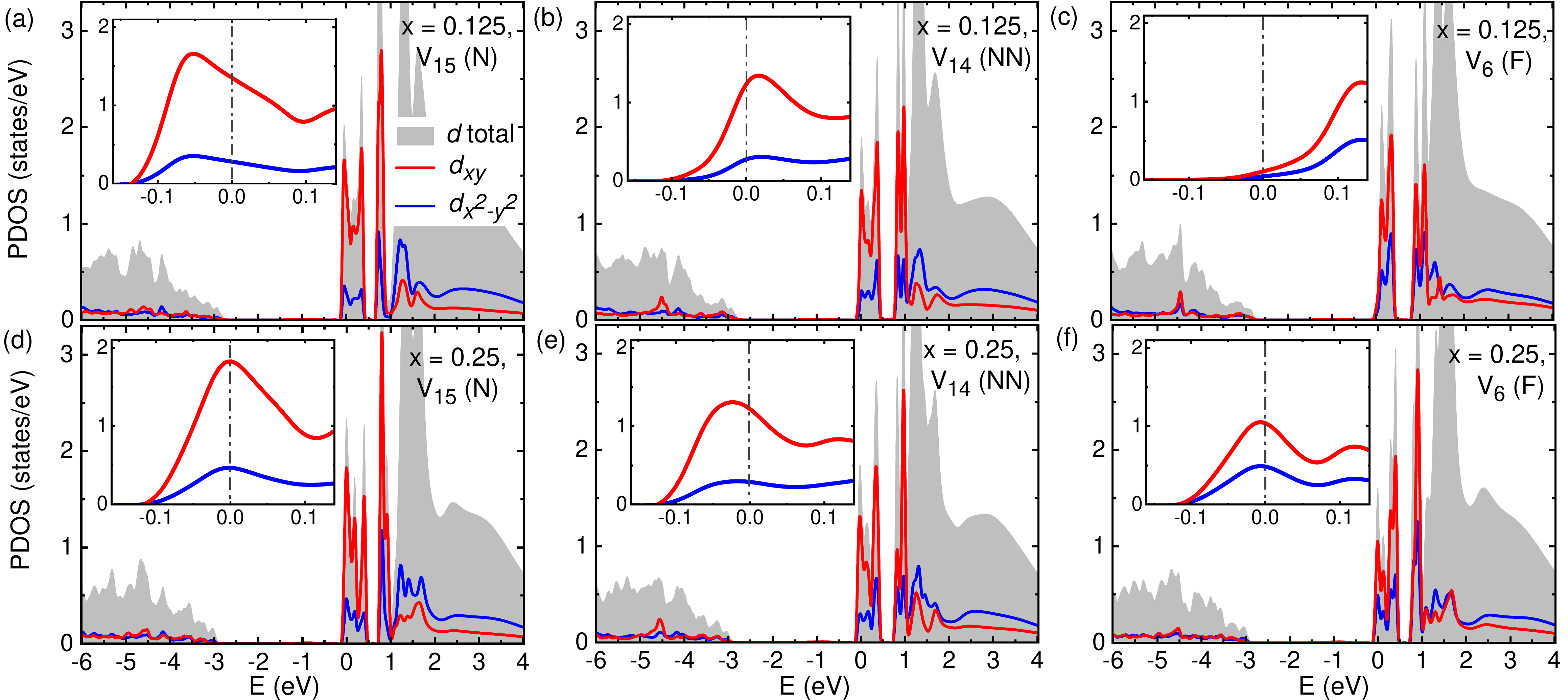}
\caption{ Projected Density of States (PDOS) onto V $d$ orbital within DMFT, for (a)-(c) 
$\alpha$-Li$_{0.125}$V$_2$O$_5$ and (d)-(f) $\alpha$-Li$_{0.25}$V$_2$O$_5$.
We choose V sites which are nearest neighbor (N) of Li (V$_{15}$), 
next nearest neighbor (NN) of Li (V$_{14}$), and far (F) from Li ion (V$_6$).
Here, we use $U = 5.5$ eV, $J = 0.5$ eV, $\lambda = 0.4$ and temperature $T = 300$ K. 
Fermi level is set to be zero, and the insets zoom in the region near the Fermi level. 
This figures show the bound polaronic effects in Li-doped V$_2$O$_5$.
 \label{fig8}}  
\end{figure*}

To explore the polaronic effect suggested in the previous experiments  \cite{Sanchez_1984, Pecquenard-1996}, 
we performed DFT+DMFT calculations using parameters analogous to those of pristine V$_2$O$_5$ 
($U = 5.5$ eV, $J = 0.5$ eV and $\lambda = 0.4$ at $T = 300$ K).
We examine two different types of structural distortions, as stated in Sec.~\ref{DMFT_Method}.
By considering two different structures, we aim to disentangle the effect of 
structural distortion and electron correlation on the electron distribution.
Surprisingly, as shown in Table~\ref{tab:polaron}, the bound polaron is more stable
than the free polaron within DMFT, while the energy difference is only 0.11 eV. 
We emphasize that the experiments observed simultaneously both types of polarons 
\cite{Sanchez_1984, Pecquenard-1996, Ioffe-PSS1970}, which are fruitfully interpreted by 
smaller energy difference between the two states within our DFT+DMFT results (Table~\ref{tab:polaron}). 
Furthermore, both states can also explain the diminishing effect of $d_{xy}$ (or $d_{xy}$ occupation)
and inhomogeneous electron distribution by scanning transmission X-ray microscopy (STXM) 
in Li-doped V$_2$O$_5$ \cite{Jesus-NatureCom-2016, Jarry-CM-2020}.


For the free polaron within DFT+DMFT, the doped electron predominantly 
occupies the V$_{15}$ site, similar to DFT+$U$.
The electron gain of V$_{15}$ is 0.46 $e$, while $N_d$ for V$_2$, V$_4$, V$_{14}$, 
and V$_{16}$ exhibit a slight increase, ranging from 0.05 to 0.08 $e$
[see Fig.~\ref{fig:LixV2O5-Nd}(a)].
The bound polaron depicted by DFT+DMFT is notably intriguing.
This state has no defect level is in the band gap and is hidden above the CBM 
[Figs.~\ref{fig8}(a)--(c)].
The electron at the Li defect level is thus empty, and doped electrons are occupying  
the V $d_{xy}$ state of the split-off band.
V ions at the lower layer (V$_{1}$, V$_{3}$, V$_{13}$, and V$_{15}$) gain  
the largest amount of electron around 0.12$-$0.16 $e$ per V atom 
[Figure \ref{fig8}(a)], 
while the upper-layer V ions (V$_{2}$, V$_{4}$, V$_{14}$, and V$_{16}$) 
also increase about 0.05 $e$ 
 [Figure \ref{fig8}(b)].
On the other hand, the other V ions far from Li ion lose 0.03 $e$ per V atom,
and their $d$ bands  are empty [Figure \ref{fig8}(c)]. 
 Our DMFT result indicates that the electron localizing in V ions close to Li site,
consistent with the experiments \cite{Sanchez_1984, Pecquenard-1996}.

To understand the reason why we cannot observe the bound polaron by DFT+U with full atomic relaxation, 
we carried our a simple test by taking the bound polaron structure within DMFT caluclation 
and then adopting the spin-polarized DFT+$U$(=4 eV)+vdW.  
As presented in Table~\ref{tab:polaron}, the total energy of the bound polaron is higher than 
the free one by 0.51 eV, indicating that bound polaron is unstable within DFT+$U$.
Our results suggest that electron localization in V$_2$O$_5$ is overestimated 
within DFT+$U$, and the discrepancy is rectified by DFT+DMFT.

Another important feature from our DMFT results is nonzero CB gap of 0.4 eV 
for Li$_{0.125}$V$_2$O$_5$, as depicted in Figures \ref{fig8}(a)-(c). 
Recent photoluminescence, optical absorption, and photoemission spectroscopy suggested that 
CB gap is ~ 0.5 eV for Li$_x$V$_2$O$_5$ ($0 \leq x \leq 1$) \cite{Wang-PRB-2016}.
Therefore, while DFT+$U$ fails to obtain nonzero CB gap for Li-doped V$_2$O$_5$,
the splitting of the split-off band and the main CB gap is successfully described by DMFT.

\subsubsection{ $\alpha$-Li$_{0.25}$V$_2$O$_5$}
\label{sec:Li0.25V2O5}

 We  now consider Li$_{0.25}$V$_2$O$_5$ by adding two Li atom in the 
1$\times$2$\times$2 supercell (the stoichiometric formula is Li$_2$V$_{16}$O$_{40}$).
From the experiment, $\alpha$ and $\epsilon$ phases are coexistent for $x$ = 0.25
\cite{COCCIANTELLI1991103, Murphy-IC-1979,Hadjead-CM-2006},
but we only consider $\alpha$-Li$_{0.25}$V$_2$O$_5$.
As shown in  
Figure~\ref{fig:LixV2O5_atmstr}, we choose two Li ions positioned at 
two of four large holes: (i) \emph{near-hole} with Li-Li distance ($d_{\mathrm{Li-Li}}$) 
of 3.166~\AA, and (ii) \emph{far-hole} with $d_{\mathrm{Li-Li}}$ = 8.104~\AA.
The far-hole configuration is more stable than the near-hole structure by 0.19 eV because the Coulomb interaction between Li$^+$ ions becomes weaker.
Hereafter we only focus on the far-hole configuration.


The electronic properties of $x = 0.25$ case behave relatively similar to the $x = 0.125$ one.
Within DFT+$U$, the electron or free polaron is strongly localized at the defect level,
i.e., two doped electrons are trapped on V$_7$ and V$_{14}$ [Table \ref{table:LixV2O5-dft} and figures \ref{fig:1Li-DFT-dos}(c)--(d)].
Therefore, the spin-up defect levels are almost degenerate
near 0.62 eV above VBM and 
[Figure \ref{fig:1Li-DFT-dos}(c)].

The bound polaronic state in $x = 0.25$ by DFT+DMFT shows that the electron is more delocalized than $x = 0.125$. 
As presented in  figures \ref{fig:LixV2O5-Nd}(b) and \ref{fig8}(d)-(f), 
$d$ bands of all V atoms gain electron and partially occupy, so the distribution of doped electrons in Li$_{0.25}$V$_2$O$_5$ become more homogeneous than Li$_{0.125}$V$_2$O$_5$.
No the defect level is occurred in the band gap, and electrons are occupied V $d_{xy}$ bands, 
which results in the metallic state compared with the prediction of insulating case by DFT+$U$.

When $x$ is increased from 0.125 to 0.25, the Fermi level within DMFT is increased since
the additional electrons are occupying the lowest conduction band ($d_{xy}$), 
while the electrons are occupying the mid-gap defect state and thus the Fermi level is unchanged within DFT+$U$. 
As shown in Figures \ref{figure4} and \ref{fig8}, the Fermi level from DMFT with respect to the 
valence band maximum (VBM) is increased from 2.52 to 2.83 eV for $x$ = 0 to 0.25.
Indeed, the photoluminescence, optical absorption and depth-resolved cathodoluminescence spectroscopies  
suggested the occurrence of the Burstein-Moss effect in Li-doped V$_2$O$_5$ 
\cite{ Wang-PRB-2016, Jarry-CM-2020, Talledo-JAP-1995}, consistent with our DMFT results.

\section{\label{conclusion} Conclusion}

Based on the DFT+$U$ and DFT+DMFT study, we have shown that the precise description 
of the electron correlation is important on the electronic structure of V$_2$O$_5$ 
and Li$_x$V$_2$O$_5$ ($x$ = 0.125 and 0.25).
For the pristine V$_2$O$_5$, we compare three experimentally quantities:
(i) band gap ($E_{\mathrm{gap}}$ = 2.6 eV), (ii) gap in the conduction band (CB gap = 0.4$-$0.5 eV), 
and the number of $d$ electrons of V ($N_d$ = 2.0).
While both experimental band gap and CB gap can be obtained using DFT+$U$,
$N_d$ is twice as large as the experimental value, indicating that the O $p$-V $d$ hybridization is 
overestimated by DFT+$U$. 

Our DMFT results shows that for the zero double counting correction, the band gap 
is not very sensitive on $U$, and it is much smaller than the experimental value 
even with $U$ = 6.5 eV.
We found that using nonzero double counting term enlarges the band gap and provides experimental value. 
Since nonzero double counting term suppress the $p$-$d$ hybridization, 
it is important on the band gap of the charge-transfer insulator V$_2$O$_5$.

The difference between DFT+$U$ and DMFT results are more dramatic for Li-doped V$_2$O$_5$.
For both Li$_{0.125}$V$_2$O$_5$ and Li$_{0.25}$V$_2$O$_5$ using DFT+$U$ method, 
only the free polaron is preferable, i.e.,  
defect levels are formed in the middle of the band gap. 
The spin-up defect levels are fully occupied by electrons from Li, 
and the conduction band is empty. 
Spatially, the electron at the defect level is localized on one vanadium site.
Our DMFT results show both types of polarons as ESP and ENDOR suggested
\cite{Pecquenard-1996, Sanchez_1984}, 
and the delocalized polaronic state is energetically more stable than the free one.
In this state, the doped electrons are localized mostly over four Vanadium sites.
Thus, the defect level is empty and hidden in the conduction band,
and the electron is occupying at the split-off band, resulting in the increase of the conduction band.
The Fermi level shift with Li doping is consistent with recently observed Burstein-Moss shift,
which is absorption energy shifts to higher energies, from the optical absorption and 
photoemission spectroscopy \cite{ Wang-PRB-2016, Jarry-CM-2020}.

\section{Acknowledgement}

This work was supported by the Assistant Secretary for Energy Efficiency and Renewable Energy,
Office of Vehicle Technologies of the US Department of Energy, through the Battery Materials Research
(BMR) program. We also acknowledge financial support from the US Department of Energy, Office of Science, Office of Basic Energy Sciences, Materials Science and Engineering Division.
We gratefully acknowledge the computing resources provided on Bebop, a high-
performance computing cluster operated by the Laboratory Computing Resource Center at Argonne
National Laboratory.

\bibliography{apssamp}

\appendix

\section{Band structure of pristine $\alpha$-V$_2$O$_5$}
\label{appendix:band}

\begin{figure}
 \centering
\includegraphics[scale=0.33]{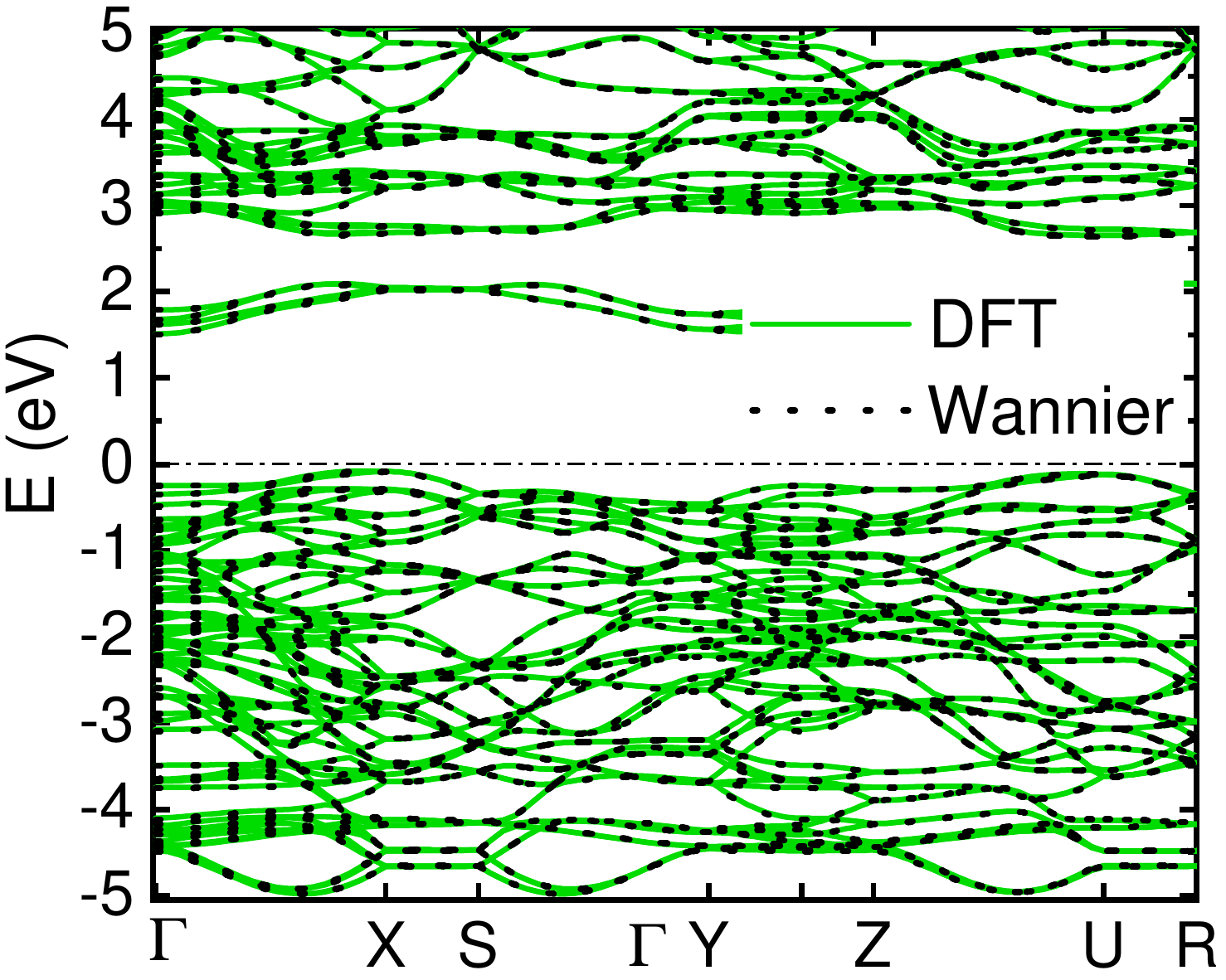}
\caption{Band structure calculated by DFT and Wannier function for pristine $\alpha$-V$_2$O$_5$. Fermi level is set at zero.
\label{appendix1}}
\end{figure}

In Figure \ref{appendix1}, we compare the DFT and Wannier band structures. 
This result indicates the plane wave functions fits very well with localized orbital ones.

\section{Optimizing Li atom position in V$_2$O$_5$ system}
\label{appendix:Li1}
\begin{figure}
 \centering
\includegraphics[scale=0.38]{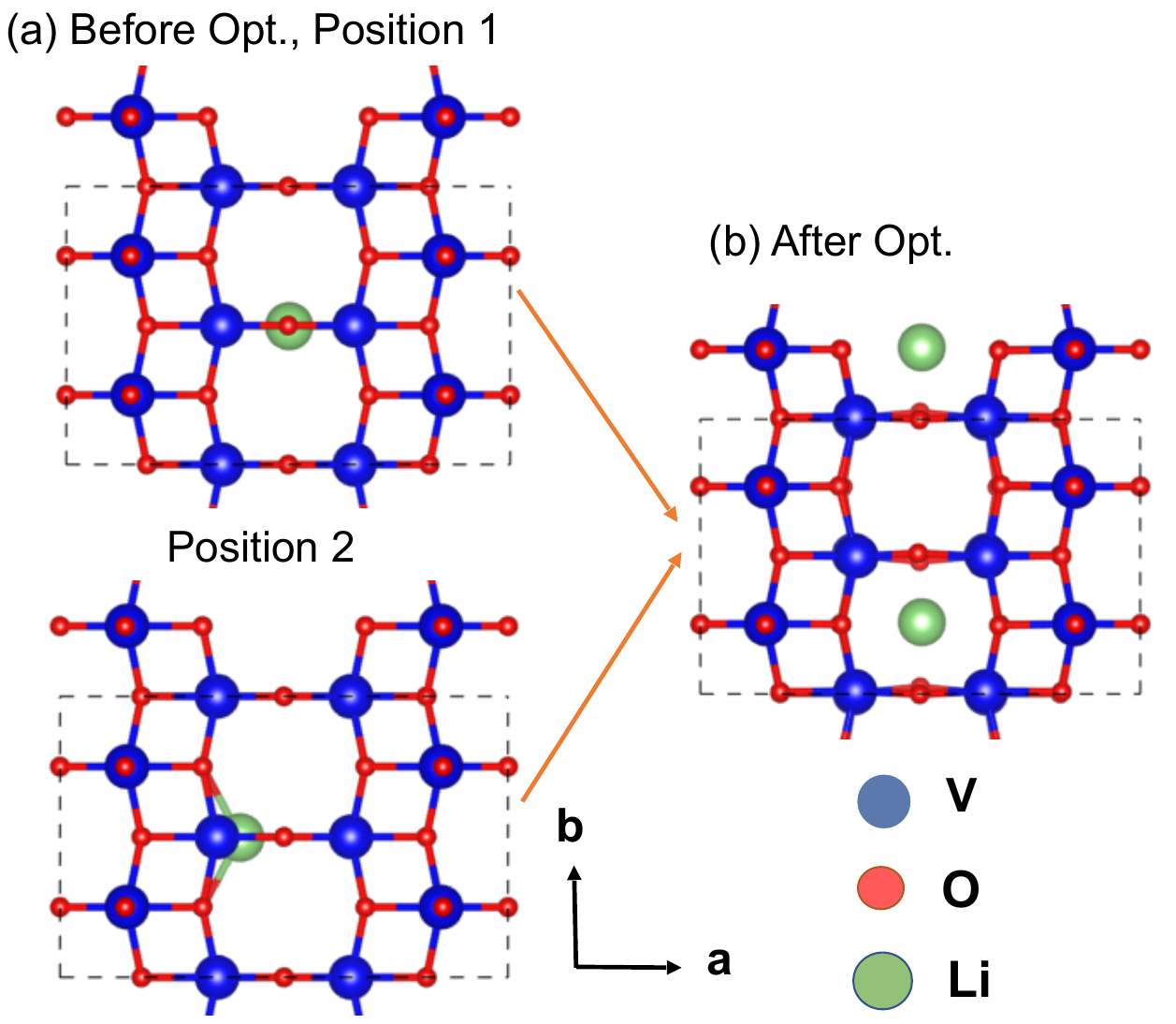}
\caption{ Atomic structure of Li$_{0.125}$V$_2$O$_5$ (a) before and (b) after optimizations.  \label{fig:appendix_1} }
\end{figure}

In this appendix, we present our relaxations of Li$_x$V$_2$O$_5$ ($x = 0.125$ and 0.25) systems. 
Several positions of Li atom in V$_2$O$_5$ framework were checked carefully. 
With $x = 0.125$, a single Li atom was inserted in 1$\times$2$\times$2 supercell, 
(stoichiometric formula of Li$_1$V$_{16}$O$_{40}$). 
First, we inserted a random position of the Li atom 
[as shown in Figure~\ref{fig:appendix_1}(a)], 
and after the relaxing process, 
it moved and located at the middle of the ``\emph{hole}'', which is  
surrounded by four vanadium atoms (front view). 
Second, in order to confirm the middle of hole is the most stable position,
we adjusted the Li$^+$ ion around it. 
We notify that the position of Li$^+$ ion in the off-center gave us a little lower energy of 20 meV than the center. 
Also, comparing to the previous DFT+$U$ results about Li-inserted V$_2$O$_5$
\cite{Scanlon-JPCC-2008, Jesus-NatureCom-2016}, 
we conclude that the most stable location for Li$^+$ ion is the the middle of the hole. 

\begin{table}
        \caption{Different configuration of Li$^+$ ion in supercell 122 using DFT+$U$, $U = 4$ eV and $J = 0.0$ eV.
            \label{tab:my_label}}
    \begin{ruledtabular}
    \begin{tabular}{c c c}
         Config. & Change E (eV)  & Mom. ($\mu_{\textrm{B}}$) \\
         \hline
         Center & 0.02 &  1
         \\
         Off-center & 0 & 1 
    \end{tabular}
\end{ruledtabular}
\end{table}

 \begin{figure}
 \centering
\includegraphics[scale=0.365]{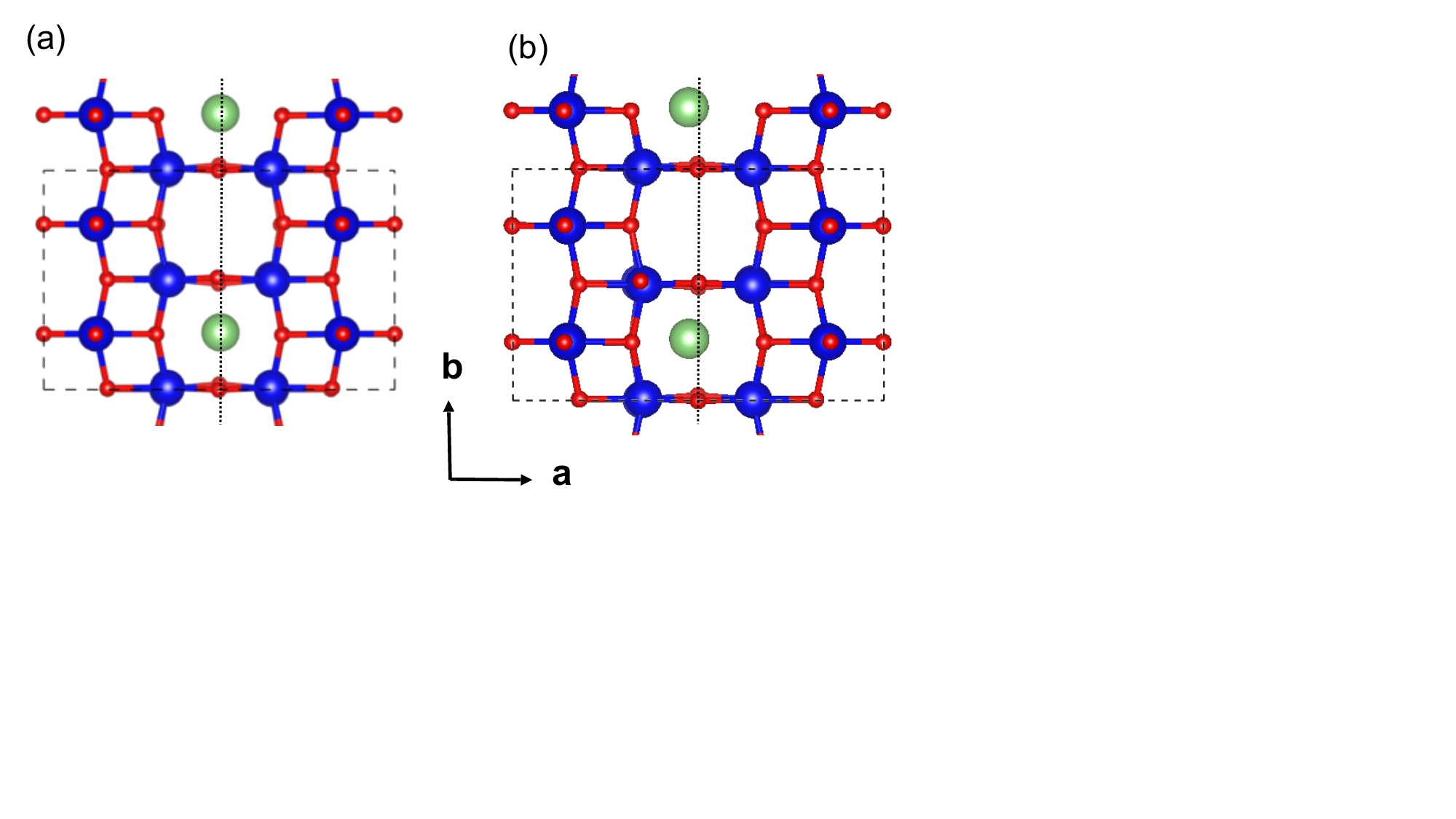}
\caption{ Atomic structure of Li$_{0.125}$V$_2$O$_5$ for (a) Li$^+$ ion at the center of the \emph{hole} 
and (b) Li$^+$ ion at the off-center of the \emph{hole}.  \label{fig:appendix2}}
\end{figure} 

\begin{figure}
 \centering
\includegraphics[scale=0.28]{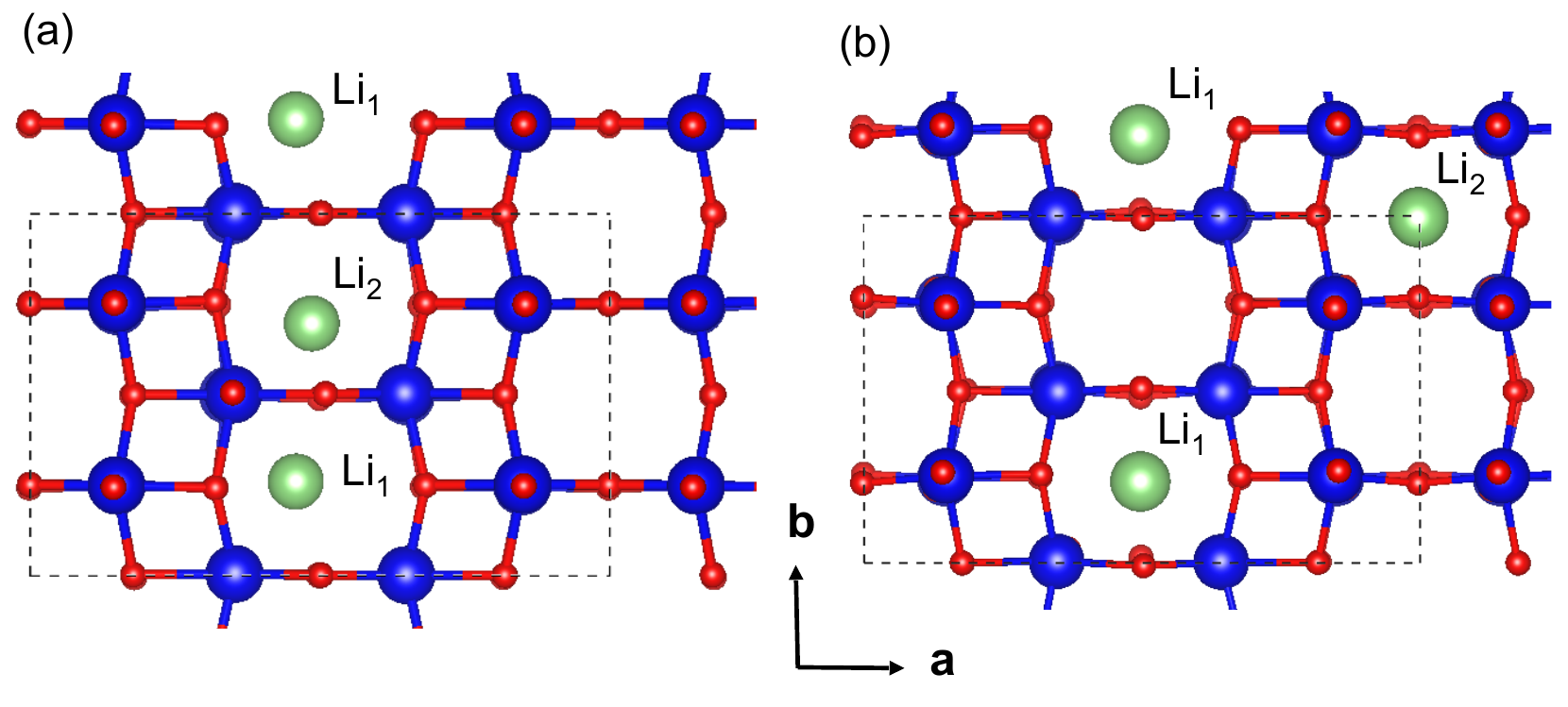}
\caption{ Atomic structure of Li$_{0.25}$V$_2$O$_5$ for (a) \emph{near-hole} 
and (b) \emph{far-hole}. }
\label{fig:Appendix3}
\end{figure}

Since the middle of hole is the most stable location for Li$^+$ ion,
with $x = 0.25$ case, there are four holes in the supercell.
So, we placed the first Li$_1 ^+$ ion in the hole, 
which is similar to $x = 0.125$ case, 
and then chose the second one in the near or far hole [Fig.~\ref{fig:Appendix3}].
As shown in Table~\ref{tab:my_label2}, we observe that 
the far-hole situation has lower energy than 
the near one by 190 meV 
by minimizing the Coulomb interaction between two Li$^{+}$ ions in the system.
We took the far-hole structure for further DMFT study.

\begin{table}
        \caption{Relative energies of the two different atomic configurations 
        for Li$_{0.25}$V$_{2}$O$_{5}$, with different spin order. 
        Here we used DFT+$U$ with $U = 4$ eV and $J = 0$ eV.
            \label{tab:my_label2}}
    \begin{ruledtabular}
    \begin{tabular}{c c c}
         Config. & Energy (eV)  & Mag. mom. ($\mu_{\textrm{B}})$ \\
         \hline
         Near-hole & 0.19
 &  2 (FM)
         \\
         Near-hole & 0.19 &  0 (AFM)
         \\
         Far-hole & 0 & 2 (FM) 
    \end{tabular}
\end{ruledtabular}
\end{table}

\section{Testing the free and bound polarons in Li$_{0.125}$V$_2$O$_5$ using DFT+$U$ and DMFT methods}
\label{detect_polarons}
\paragraph{Free polaron:} We used DFT+$U$(=4 eV)+vdW to optimize the atomic structure. 
From the optimal structure, we solve non-spin-polarized Kohn-Sham equation using DFT+$U$(=0 eV)+vdW within VASP.   
Then, we performed the localized orbital interpolation.  
Finally, we apply the correlation and hybridization effects within DMFT to the system. 
By that way, the free polaron was observed by DFT+DMFT. 
This is a standard procedure for DFT+DMFT calculation, 
which is described in Refs.~\cite{Singh-CPC-2021, Park-prb-2020} .
\paragraph{Bound polaron:}  

There are two reasons which we want to observe the atomic scale existence of the bound polaron. 
First, experimental measurements including ESR, ENDOR and electronic conductivity suggested the coexistence of free and bound polaron in $\alpha$-Li$_x$V$_2$O$_5$. 
Second, none of the DFT+$U$ works have predicted about this polaron. 
However, we have seen that at DFT+$U$(= 0 eV), the doped electron is more delocalizing in the system. So, we altered the standard DFT+DMFT computation as following:
(1) From the DFT+$U$(=4 eV)+vdW structure, we reoptimized it with $U = 0$ eV with a fixed lattice parameters and non-spin polarized schemes inside VASP. 
(2) We took this structure for further steps such as wannierisation and self-consistent DMFT calculation.  
We also test the boundly polaronic state in DFT+$U$ by simply applying $U = 4$ eV on the optimal DFT as shown in table~\ref{tab:polaron}.

\section{Electron localization function}
\label{ELF}
 \begin{figure}
 \centering
\includegraphics[scale=0.18]{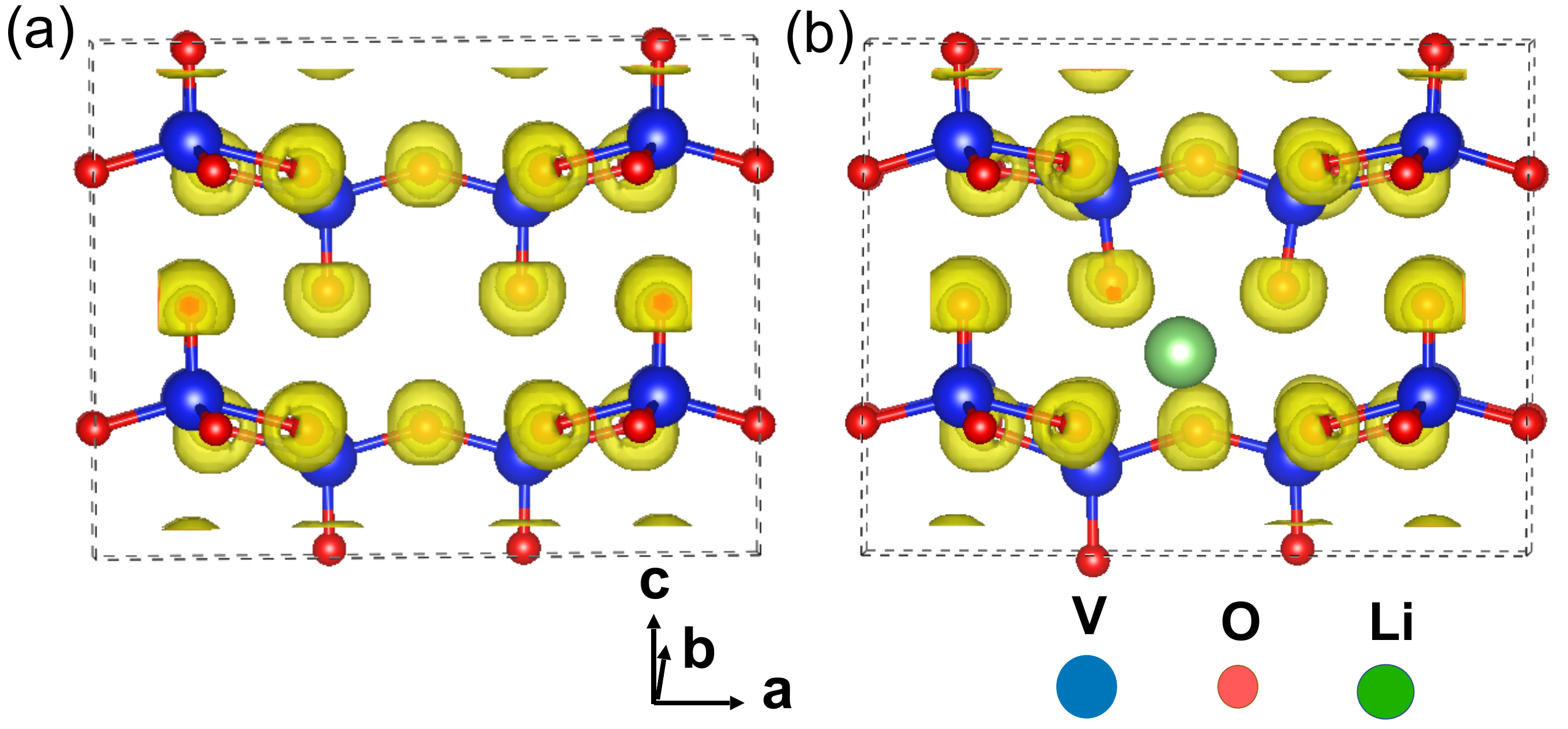}
\caption{Isosurface plots of the electron localization function for (a) V$_2$O$_5$ and (b) Li$_{0.125}$V$_2$O$_5$.  \label{fig:appendixd}}
\end{figure} 

With the optimal structures of Li$_x$V$_2$O$_5$ ($x = 0$ and 0.125) obtaining in Sec.~\ref{DFT}, 
we plot their electron localization function (ELF) isosurfaces within DFT+$U$ using $U$ = 4 eV 
(as shown in Figure~\ref{fig:appendixd}). 
In the nature bond between V and O, we recognize that electrons is localized at O sites, 
which indicates the ionic bonding \cite{ELF-1, ELF-2, ELF-3}.

\nocite{apsrev42Control}
\end{document}